\documentclass{article}
\usepackage{amssymb}
\usepackage{amsmath}

\begin{document}

\noindent \textbf{The Electromagnetic field equations for moving
media\bigskip \medskip }

\noindent T Ivezi\'{c}

\noindent Ru%
\mbox
{\it{d}\hspace{-.15em}\rule[1.25ex]{.2em}{.04ex}\hspace{-.05em}}er Bo\v{s}%
kovi\'{c} Institute, P.O.B. 180, 10002 Zagreb, Croatia

\noindent E-mail: ivezic@irb.hr\textbf{\bigskip \bigskip }

In this paper a formulation of the field equation for moving media is
developed by the generalization of an axiomatic geometric formulation of the
electromagnetism in vacuum (Ivezi\'{c} T 2005 \textit{Found.\ Phys. Lett.}
\textbf{18} 401). First, the field equations with bivectors $F(x)$ and $%
\mathcal{M}(x)$ are presented and then these equations are written with the
4D vectors $E(x)$, $B(x)$, $P(x)$ and $M(x)$. The latter contain both the 4D
velocity vector $u$ of a moving medium and the 4D velocity vector $v$ of the
observers who measure $E$ and $B$ fields. They do not appear in previous
literature. All these equations are also written in the standard basis and
compared with Maxwell's equations with 3D vectors. In this approach the Amp%
\`{e}re-Maxwell law and Gauss's law are inseparably connected in one law and
the same happens with Faraday's law and the law that expresses the absence
of magnetic charge. It is shown that Maxwell's equations with 3D vectors and
the field equations with 4D geometric quantities are not equivalent in 4D
spacetime\bigskip \medskip

\bigskip \bigskip

\noindent \textbf{1. Introduction\bigskip }

\noindent The field equations for moving media in a relativistically
covariant formulation were first presented by Minkowski [1]. An axiomatic
geometric formulation of electromagnetism in vacuum is presented in [2].
That formulation is with only one axiom: the field equation for the bivector
field $F$. In this paper the formulation from [2] is generalized to moving
media. The geometric approach to special relativity (SR) that is used in
[2], and in this paper as well, exclusively deals either with the abstract,
coordinate-free, four-dimensional (4D) geometric quantities, e.g., vectors
(4-vectors in the usual notation) $E(x)$, $B(x)$, .. ($x$ is the position
vector), or with their representations in some basis, the 4D
coordinate-based geometric quantities (CBGQs) comprising \emph{both}
components and \emph{a basis}, e.g., $E=E^{\nu }\gamma _{\nu }$. In that
approach, which is called \textquotedblleft invariant special
relativity\textquotedblright\ (ISR), an independent physical reality is
attributed to 4D geometric quantities and not, as usual, to 3D quantities.
Every 4D CBGQ is \emph{invariant} under the passive Lorentz transformations
(LT); the components transform by the LT and the basis by the inverse LT
leaving the whole CBGQ unchanged. The invariance of a 4D CBGQ under the
passive LT reflects the fact that such 4D geometric quantity represents
\emph{the same physical quantity} for relatively moving inertial observers.
This invariance is the reason for the name ISR. The principle of relativity
is naturally satisfied, if the physical laws are written with the abstract
4D geometric quantities or with 4D CBGQs. There is no need to postulate it
outside the mathematical formulation of the theory as in Einstein's
formulation of SR [3]. The paper is organized as follows.

In Sec. 2 the basic field equation for moving media (Eq. (\ref{F4})) is
expressed in terms of\textbf{\ }the bivector $F=F(x)$ that represents the
electromagnetic field and the generalized magnetization-polarization
bivector $\mathcal{M=M}(x)$. Then that equation with abstract quantities
(AQs) is written with CBGQs in the standard basis. The resulting equation is
separated into two equations, the equation with sources and the equation
without sources. The field equation with sources is also written in the
\textquotedblleft source representation\textquotedblright\ both with AQs and
with CBGQs in the $\{\gamma _{\mu }\}$ basis, according to which the sources
of $F$ field are the true currents $j^{(C)}$ and the
magnetization-polarization current density $\partial \cdot \mathcal{M}$.

In Sec. 3 the decomposition of $F$ is presented. $F$ is decomposed into the
electric field vector $E$, the magnetic field vector $B$ and the velocity
vector $v$ of the observers who measure $E$ and $B$ fields (in the usual
notation $E$, $B$, $v$, ... are called 4-vectors). Similarly, $\mathcal{M}$
is decomposed into the polarization vector $P(x)$, the magnetization vector $%
M(x)$ and the bulk velocity vector $u$ of the medium. Inserting these
decompositions into the basic field equation with $F$\ and $\mathcal{M}$\
(Eq. (\ref{F4}))\ we find the general form of the field equations for a
magnetized and polarized moving medium expressed in terms of $E(x)$, $B(x)$,
$P(x)$ and $M(x)$, which are named the field equations in the Amp\`{e}rian
form. In equation with the geometric product (Eq. (\ref{F7})), i.e., in its
vector part (Eq. (\ref{A1})), there are two different velocities $u$ and $v$
and, as I am aware, these field equations do not appear in previous
literature. They are important results that are obtained in this paper. All
AQs $E$, $B$, $P$ and $M$ are represented in the standard basis $\left\{
\gamma _{\mu }\right\} $ in order to compare these basic field equations
with usual formulations that deal with 3D vectors. The equation (\ref{A1})
is also written in the \textquotedblleft source
representation\textquotedblright\ both with AQs and with CBGQs. It is
visible from that representation that the sources of $E$ and $B$ fields are
the true current density $j^{(C)}$ and the $P$ and $M$ vectors. \emph{The
field equations with AQs} (Eq. (\ref{F4})) \emph{and} (Eq. (\ref{F7})) \emph{%
or the corresponding equations with CBGQs} \emph{comprise and generalize all
usual Maxwell's equations (with 3D vectors) for moving media.}

In Sec. 4 we first present a brief review of the existence of the
fundamental difference between the usual transformations (UT) of the
electric and magnetic fields as 3D vectors and the mathematically correct LT
of 4D geometric quantities that represent the electric and magnetic fields
in 4D spacetime. Then, the basic field equations with CBGQs are compared
with usual Maxwell's equations with 3D vectors for the case when the
observers are at rest in a stationary medium. It is shown that these two
formulations are not equivalent, because the UT are not the LT.

In Sec. 5 the similar comparison is presented for the case when the
observers are at rest in the laboratory frame, but material medium is
moving. Again the same result is obtained as in Sec. 4 that these two
formulations are not equivalent since 3D quantities do not properly
transform under the LT.

In Sec. 6 the comparison with Galilean Electromagnetism is presented. It is
shown that both Galilean limits are ill-defined in 4D spacetime and that
they are not a quasi-static approximation of the relativistically correct
field equations.

In Secs. 7, 7.1 and 7.2 the motional emf $\varepsilon $ is calculated using
3D quantities and their UT, Sec. 7.1, and 4D geometric quantities and their
mathematically correct LT, Sec. 7.2. In Sec. 7.1 it is shown that if $%
\varepsilon $ is defined by the usual definition with 3D quantities then $%
\varepsilon $ is different in relatively moving inertial frames, $%
\varepsilon =UBl$ in $S$ and $\varepsilon ^{\prime }=\gamma UBl$ in $%
S^{\prime }$, which means that the principle of relativity is not satisfied
in the usual formulation of electromagnetism with 3D quantities and their UT
of $\mathbf{E}$ and $\mathbf{B}$ (and the UT of $\mathbf{P}$ and $\mathbf{M}$%
). In Sec. 7.2 it is shown that if $\varepsilon $ is defined as an invariant
4D quantity, the Lorentz scalar, then always the same value for $\varepsilon
$ is obtained, $\varepsilon =\gamma UBl$. This result unambiguously shows
that the principle of relativity is naturally satisfied in the approach with
4D geometric quantities and their LT.

In Sec. 8 the discussion of the results and the conclusions are
presented.\bigskip \bigskip

\noindent \textbf{2. The basic field equation for moving media in terms of }$%
F$ and $\mathcal{M}$\bigskip

\noindent In this paper, the geometric algebra formalism [4] will be used.
In order to compare Eq. (\ref{F7}), i.e., Eqs. (\ref{A1}) and (\ref{A2}),
with the usual formulations of electromagnetism with 3D vectors we shall
represent all abstract quantities in (\ref{F7}), i.e., (\ref{A1}) and (\ref%
{A2}), with 4D CBGQs in the standard basis $\left\{ \gamma _{\mu }\right\} $%
. Thus, for the reader's convenience, all equations will be written not only
with the abstract multivectors but also with CBGQs in the standard basis.
Therefore, \emph{the knowledge of the geometric algebra is not required for
the understanding of this presentation.}

However, a very brief summary of the geometric algebra will be provided
here. The geometric (Clifford) product is written by simply juxtaposing
multivectors $AB$. The geometric product of a grade-$r$ multivector $A_{r}$
with a grade-$s$ multivector $B_{s}$ decomposes into $A_{r}B_{s}=\left%
\langle AB\right\rangle _{\ r+s}+\left\langle AB\right\rangle _{\
r+s-2}...+\left\langle AB\right\rangle _{\ \left\vert r-s\right\vert }$. The
inner and outer (or exterior) products are the lowest-grade and the
highest-grade terms respectively of the above series; $A_{r}\cdot
B_{s}\equiv \left\langle AB\right\rangle _{\ \left\vert r-s\right\vert }$
and $A_{r}\wedge B_{s}\equiv \left\langle AB\right\rangle _{\ r+s}$. For
vectors $a$ and $b$ we have: $ab=a\cdot b+a\wedge b$, where $a\cdot b\equiv
(1/2)(ab+ba)$, $a\wedge b\equiv (1/2)(ab-ba)$. Usually the above mentioned
standard basis is introduced. The generators of the spacetime algebra (the
Clifford algebra generated by Minkowski spacetime) are taken to be four
basis vectors $\left\{ \gamma _{\mu }\right\} ,\mu =0,...,3,$ satisfying $%
\gamma _{\mu }\cdot \gamma _{\nu }=\eta _{\mu \nu }=diag(+---).$ The basis
vectors $\gamma _{\mu }$ generate by multiplication a complete basis for the
spacetime algebra: $1$, $\gamma _{\mu }$, $\gamma _{\mu }\wedge \gamma _{\nu
}$, $\gamma _{\mu }\gamma _{5}$, $\gamma _{5}$ ($2^{4}=16$ independent
elements). $\gamma _{5}$ is the right-handed unit pseudoscalar, $\gamma
_{5}=\gamma _{0}\wedge \gamma _{1}\wedge \gamma _{2}\wedge \gamma _{3}$. Any
multivector can be expressed as a linear combination of these 16 basis
elements of the spacetime algebra.

This basis, the standard basis $\left\{ \gamma _{\mu }\right\} $, is a
right-handed orthonormal frame of vectors in the Minkowski spacetime $M^{4}$
with $\gamma _{0}$ in the forward light cone, $\gamma _{0}^{2}=1$ and $%
\gamma _{k}^{2}=-1$ ($k=1,2,3$). The $\left\{ \gamma _{\mu }\right\} $ basis
corresponds to Einstein's system of coordinates in which the Einstein
synchronization of distant clocks [3] and Cartesian space coordinates $x^{i}$
are used in the chosen inertial frame of reference.

The field equation in vacuum in the geometric algebra formalism is:

\begin{equation}
\partial F=j/\varepsilon _{0}c,\quad \partial \cdot F+\partial \wedge
F=j/\varepsilon _{0}c,  \label{F1}
\end{equation}%
i.e., with the CBGQs in the $\{\gamma _{\mu }\}$ basis, that equation becomes

\begin{equation}
\partial _{\alpha }F^{\alpha \beta }\gamma _{\beta }-\partial _{\alpha }\
^{\ast }F^{\alpha \beta }\gamma _{5}\gamma _{\beta }=(1/\varepsilon
_{0}c)j^{\beta }\gamma _{\beta },  \label{c1}
\end{equation}%
where $^{\ast }F^{\alpha \beta }=(1/2)\varepsilon ^{\alpha \beta \gamma
\delta }F_{\gamma \delta }$ is the usual dual tensor, $\varepsilon ^{\alpha
\beta \gamma \delta }$ is the totally skew-symmetric Levi-Civita
pseudotensor. The usual covariant form of Eq. (\ref{c1}), i.e., \emph{only
the basis components in the} $\left\{ \gamma _{\mu }\right\} $ \emph{basis,}
are two equations, the equation with sources $\partial _{\alpha }F^{a\beta
}=j^{\beta }/\varepsilon _{0}c$, and that one without sources $\partial
_{\alpha }\ ^{\ast }F^{\alpha \beta }=0$. It is shown in [2] that the
bivector $F=F(x)$, which represent the electromagnetic field, can be taken
as the primary quantity for electromagnetism and the field equation for $F$,
Eq. (\ref{F1}), is the basic equation. As shown in [2], the bivector field $%
F $ yields a complete description of the electromagnetic field and, in fact,
there is no need to introduce either the field vectors or the potentials.
For the given sources the Clifford algebra formalism enables one to find in
a simple way the electromagnetic field $F$, see Eqs. (7) and (8) in [2].

If $j$ is the total current density then (\ref{F1}), i.e., (\ref{c1}), \emph{%
holds unchanged in moving medium as well.} The equation (\ref{F1}) can be
separated into the field equation with sources and that one without sources
as
\begin{equation}
\partial \cdot F=j/\varepsilon _{0}c,\quad \partial \wedge F=0,  \label{R1}
\end{equation}%
i.e., with CBGQs in the $\{\gamma _{\mu }\}$ basis they are%
\begin{equation}
\partial _{\alpha }F^{\alpha \beta }\gamma _{\beta }=(1/\varepsilon
_{0}c)j^{\beta }\gamma _{\beta },\quad \partial _{\alpha }\ ^{\ast
}F^{\alpha \beta }\gamma _{5}\gamma _{\beta }=0.  \label{fvc}
\end{equation}%
Since $j$ is a vector the trivector part is identically zero (in the absence
of a magnetic charge).

The total current density vector $j$ can be decomposed as

\begin{equation}
j=j^{(C)}+j^{(\mathcal{M})},  \label{F2}
\end{equation}%
where $j^{(C)}$ is the conduction current density of the \emph{free} charges
and $j^{(M)}$ is the magnetization-polarization current density of the \emph{%
bound} charges

\begin{equation}
j^{(\mathcal{M})}=-c\partial \mathcal{M}=-c\partial \cdot \mathcal{M}
\label{F3}
\end{equation}%
($\partial \wedge \mathcal{M}=0$, since $j^{(\mathcal{M})}$ is a vector). $%
\mathcal{M}$ is the generalized magnetization-polarization bivector $%
\mathcal{M=M}(x)$.

Then (\ref{F1}) (i.e., (\ref{R1})) can be written as%
\begin{equation}
\partial (\varepsilon _{0}F+\mathcal{M})=j^{(C)}/c;\quad \partial \cdot
(\varepsilon _{0}F+\mathcal{M})=j^{(C)}/c,\ \partial \wedge F=0.  \label{F4}
\end{equation}%
The trivector part, i.e., the field equation without sources, remained
unchanged, because it is not affected by the separation of the current
density vector $j$ into free and bound parts; that part does not contain $j$%
. The equations (\ref{F4}) are the primary equations for electromagnetism in
moving media. In most materials $\mathcal{M}$ is a function of the field $F$
and this dependence is determined by the constitutive relations. Recently,
they are discussed in detail in [5]. In that case (\ref{F4}) are
well-defined equations for $F$. The constitutive relations from [5] with 4D
geometric quantities, which correctly transform under the LT, are compared
with Minkowski's constitutive relations with 3D vectors and several
essential differences are pointed out. They are caused by the fact that, as
in the case with the field equations that are investigated here, the UT of
3D vectors $\mathbf{E}$, $\mathbf{B}$, $\mathbf{P}$, $\mathbf{M}$, etc. are
not the LT. Furthermore, in [5], the physical explanation is presented for
the existence of the magnetoelectric effect in moving media that essentially
differs from the traditional explanation.

If (\ref{F4}) is written with CBGQs in the standard basis it becomes
\begin{equation}
\partial _{\alpha }(\varepsilon _{0}F^{\alpha \beta }+\mathcal{M}^{\alpha
\beta })\gamma _{\beta }-\partial _{\alpha }(\varepsilon _{0}\ ^{\ast
}F^{\alpha \beta })\gamma _{5}\gamma _{\beta }=c^{-1}j^{(C)\beta }\gamma
_{\beta },  \label{fcb}
\end{equation}%
which can be separated into two equations, the equation with sources%
\begin{equation}
\partial _{\alpha }(\varepsilon _{0}F^{\alpha \beta }+\mathcal{M}^{\alpha
\beta })\gamma _{\beta }=c^{-1}j^{(C)\beta }\gamma _{\beta }  \label{fs}
\end{equation}%
and the equation without sources, which is the same as in the vacuum%
\begin{equation}
\partial _{\alpha }\ ^{\ast }F^{\alpha \beta }\gamma _{5}\gamma _{\beta }=0.
\label{fws}
\end{equation}

Instead of dealing with the axiomatic formulation of electromagnetism for
moving media that uses only the \emph{local }form of the field equation%
\textbf{\ }(\ref{F4}) one can construct the equivalent integral form simply
replacing $F$ by $F+\mathcal{M}/\varepsilon _{0}$ in Eqs. (18), (21), (22)
and also $j$ by $j^{(C)}$ in (21), (22) in [2]. However, the integral form
will not be investigated here.

Proceeding in the same way as in [2] one can derive from (\ref{F4}) the
stress-energy vector $T(n)$ for a moving medium simply replacing $F$ by $F+%
\mathcal{M}/\varepsilon _{0}$ in Eqs. (26), (37-47) in [2]. For example, Eq.
(26) in [2] becomes
\begin{equation}
T(n)=T(n(x),x)=-(\varepsilon _{0}/2)\left\langle (F+\mathcal{M}/\varepsilon
_{0})n(F+\mathcal{M}/\varepsilon _{0})\right\rangle _{1}.  \label{t1}
\end{equation}%
$T(n)$ is a vector-valued linear function on the tangent space at each
spacetime point $x$ describing the flow of energy-momentum through a
hypersurface with unit normal vector $n=n(x)$. The expression for $T(n)$, $%
T(n)=Un+(1/c)S$, Eq. (41) in [2], will remain unchanged, but the energy
density $U$ and the Poynting vector $S$ will change according to the
described replacement. All this with $T(n)$ will not be discussed in this
paper, but in a separate paper in which the Abraham-Minkowski controversy
will also be examined in a new way.

Another form of the field equation with sources (\ref{fs}) is the
\textquotedblleft source representation\textquotedblright\
\begin{equation}
\partial \cdot \varepsilon _{0}F=j^{(C)}/c-\partial \cdot \mathcal{M},
\label{Fj}
\end{equation}%
i.e., with the CBGQs in the $\{\gamma _{\mu }\}$ basis%
\begin{equation}
\partial _{\alpha }(\varepsilon _{0}F^{\alpha \beta })\gamma _{\beta
}=(c^{-1}j^{(C)\beta }-\partial _{\alpha }\mathcal{M}^{\alpha \beta })\gamma
_{\beta }  \label{Fj1}
\end{equation}%
according to which the sources of the fundamental electromagnetic field $F$
are the true currents $j^{(C)}$ and the magnetization-polarization current
density $\partial \cdot \mathcal{M}$, i.e., the space-time changes of the
generalized magnetization-polarization bivector $\mathcal{M}$.

In previous formulations of electromagnetism in media (at rest, or moving),
starting with Minkowski (his $f_{hk}$), [1], the electromagnetic excitation
tensor is introduced, see, e.g., a modern textbook on classical
electromagnetism, [6], or the papers [7-9] and references therein, in the
recent - Annalen der Physik, Special Topic Issue 9-10/2008: The Minkowski
spacetime of special relativity - 100 years after its discovery. Here, in (%
\ref{F4}), $\mathcal{H}$ can be introduced as
\begin{equation}
\mathcal{H}=\varepsilon _{0}F+\mathcal{M}.  \label{F5}
\end{equation}%
However, it is worth noting that (\ref{F5}) is in some sense unsatisfactory,
since physically different kind of entities are mixed in it; an
electromagnetic field $F$ and a matter field $\mathcal{M}$, i.e., the
magnetization-polarization bivector. Moreover, as will be seen in the next
section, in general, two different velocity vectors, $v$ - the velocity of
the observers and $u$ - the velocity of the moving medium, enter into the
decompositions of $F$ and $\mathcal{M}$, Eqs. (\ref{E2}) and (\ref{M1}),
respectively. This fact causes that \emph{the usual decomposition of} $%
\mathcal{H}$ \emph{into the electric and magnetic excitations, Eq.} (\ref{h1}%
), \emph{is not possible in the general case but only in the case if} $u=v$,
or if both decompositions (\ref{E2}) and (\ref{M1}) are made with the same
velocity vector, either $u$ or $v$. In that case $\mathcal{H}$ can be
introduced in (\ref{F4}) and the usual form of the field equations in moving
media is obtained%
\begin{equation}
\partial \cdot \mathcal{H}=j^{(C)}/c,\quad \partial \wedge F=0,  \label{F6}
\end{equation}%
i.e., with CBGQs in the $\{\gamma _{\mu }\}$ basis,%
\begin{equation}
\partial _{\alpha }\mathcal{H}^{\alpha \beta }\gamma _{\beta
}=c^{-1}j^{(C)\beta }\gamma _{\beta },\quad \partial _{\alpha }\ ^{\ast
}F^{\alpha \beta }\gamma _{5}\gamma _{\beta }=0.  \label{hf}
\end{equation}%
\bigskip \bigskip

\noindent \textbf{3. The basic field equation for moving media in terms of }$%
E$, $B$

\textbf{and} $P$, $M$\bigskip

\noindent In this paper instead of using (\ref{F5}), (\ref{F6}) and (\ref{hf}%
) we deal with (\ref{F4}), i.e., with (\ref{fcb}), as the basic field
equations. In that equation bivectors $F$ and $\mathcal{M}$ can be
decomposed. First, the decomposition of $F$ is considered. There is \emph{a
mathematical theorem according to which any antisymmetric tensor of the
second rank can be decomposed into two space-like vectors and the unit
time-like vector.} For the proof of that theorem in geometric terms see,
e.g., [10]. When applied to the bivector $F$, e.g., Eq. (13) in [2], this
yields

\begin{equation}
F=E\wedge v/c+(IcB)\cdot v/c,  \label{E2}
\end{equation}%
where the electric and magnetic fields are represented by vectors $E(x)$ and
$B(x)$. The unit pseudoscalar $I$ is defined algebraically without
introducing any reference frame as in Sec. 1.2 in [4] (Hestenes D and
Sobczyk G). We choose $I$ in such a way that when $I$ is represented in the $%
\left\{ \gamma _{\mu }\right\} $ basis it becomes $I=\gamma _{0}\wedge
\gamma _{1}\wedge \gamma _{2}\wedge \gamma _{3}=\gamma _{5}$. With such
choice for $I$, $\left\{ \gamma _{1},\gamma _{2},\gamma _{3}\right\} $ form
a right-handed orthonormal set, as usual for a 3D Cartesian frame. The LT
(boosts) do not change the orientation for spacetime. Here, in the whole
paper, under the name LT we shall only consider - boosts.

If (\ref{E2}) is written with CBGQs in the $\{\gamma _{\mu }\}$ basis it
becomes%
\begin{equation}
F=(1/2)F^{\mu \nu }\gamma _{\mu }\wedge \gamma _{\nu },\ F^{\mu \nu
}=(1/c)(E^{\mu }v^{\nu }-E^{\nu }v^{\mu })+\varepsilon ^{\mu \nu \alpha
\beta }v_{\alpha }B_{\beta },  \label{fm}
\end{equation}%
where $\gamma _{\mu }\wedge \gamma _{\nu }$ is the bivector basis. In the
same way as for any other CBGQ it holds that bivector $F$ \emph{is the same
4D quantity} for relatively moving inertial observers and for all bases
chosen by them, $F=(1/2)F^{\mu \nu }\gamma _{\mu }\wedge \gamma _{\nu
}=(1/2)F^{\prime \mu \nu }\gamma _{\mu }^{\prime }\wedge \gamma _{\nu
}^{\prime }=...$ .

Minkowski, Sec. 11.6, [1], see also [11], was the first who introduced
vectors (4-vectors in the usual notation) of the electric and magnetic
fields and the velocity vector, $\Phi $, $\Psi $ and $w$, respectively, in
his notation, and presented the decomposition of $F$, his equation (55),
that corresponds to (\ref{E2}). Note that he considered that $w$, $\Phi $
and $\Psi $ are $1\times 4$ matrices and $F$ is a $4\times 4$ matrix. Thus
he worked with components of the geometric quantities taken in the standard
basis $\left\{ \gamma _{\mu }\right\} $.

The vector $v$ in the decomposition (\ref{E2}) is interpreted as the
velocity vector of the observers who measure $E$ and $B$ fields. Then $E(x)$
and $B(x)$ are defined with respect to $v$, i.e., with respect to the
observer, as
\begin{equation}
E=F\cdot v/c,\quad B=-(1/c)I(F\wedge v/c).  \label{E1}
\end{equation}%
It also holds that $E\cdot v=B\cdot v=0$; both $E$ and $B$ are space-like
vectors. It is visible from (\ref{E1}) that $E$ \emph{and} $B$ \emph{depend
not only on} $F$ \emph{but on} $v$ \emph{as well}.

If (\ref{E1}) is written with CBGQs in the $\{\gamma _{\mu }\}$ basis it
becomes%
\begin{equation}
E=E^{\mu }\gamma _{\mu }=(1/c)F^{\mu \nu }v_{\nu }\gamma _{\mu },\quad
B=B^{\mu }\gamma _{\mu }=(1/2c^{2})\varepsilon ^{\mu \nu \alpha \beta
}F_{\nu \alpha }v_{\beta }\gamma _{\mu }.  \label{ebv}
\end{equation}%
As $F$ is antisymmetric it holds that $E^{\mu }v_{\mu }=B^{\mu }v_{\mu }=0$.
Only three components of $E$ and $B$ in any basis are independent. However,
as $E$ and $B$ depend not only on $F$ but on $v$ as well this result does
not mean that three spatial components of $E$, or $B$, are necessarily
independent components. The form of $v$ in a given inertial frame will
determine which three components are independent. The relations (\ref{E2}) -
(\ref{ebv}) are mathematically correct definitions.

Similarly, using the same theorem, the bivector $\mathcal{M(}x\mathcal{)}$
can be decomposed into two space-like vectors, the polarization vector $P(x)$
and the magnetization vector $M(x)$ and the unit time-like vector $u/c$
\begin{equation}
\mathcal{M}=P\wedge u/c+(MI)\cdot u/c^{2}.  \label{M1}
\end{equation}%
There is the rest frame for a medium, i.e., for $\mathcal{M}$, or $P$ and $M$%
, and therefore the vector $u$ in the decomposition (\ref{M1}) may be
identified with bulk velocity vector of the medium in spacetime. Integral
curves of $u$ define the averaged world-lines of identifiable constituents
of the medium. If (\ref{M1}) is written with CBGQs in the $\{\gamma _{\mu
}\} $ basis it becomes%
\begin{equation}
\mathcal{M}=(1/2)\mathcal{M}^{\mu \nu }\gamma _{\mu }\wedge \gamma _{\nu },\
\mathcal{M}^{\mu \nu }=(1/c)(P^{\mu }u^{\nu }-P^{\nu }u^{\mu
})+(1/c^{2})\varepsilon ^{\mu \nu \alpha \beta }M_{\alpha }u_{\beta }.
\label{mp}
\end{equation}%
The vectors $P(x)$ and $M(x)$ are determined by $\mathcal{M(}x\mathcal{)}$
and the unit time-like vector $u/c$ as

\begin{equation}
P=\mathcal{M}\cdot u/c,\quad M=cI(\mathcal{M}\wedge u/c)  \label{M2}
\end{equation}%
and it holds that $P\cdot u=M\cdot u=0$. As in the case with $F$, it can be
seen from (\ref{M2}) that $P$ and $M$ depend not only on $\mathcal{M}$ but
on $u$ as well. $P(x)$ and $M(x)$ from (\ref{M2}) can be written as CBGQs in
the $\{\gamma _{\mu }\}$ basis

\begin{equation}
P=(1/c)\mathcal{M}^{\mu \nu }u_{\nu }\gamma _{\mu },\quad M=(1/2)\varepsilon
^{\mu \nu \alpha \beta }\mathcal{M}_{\alpha \nu }u_{\beta }\gamma _{\mu },
\label{pm}
\end{equation}%
with $P^{\mu }u_{\mu }=M^{\mu }u_{\mu }=0$.

Usually, only the velocity vector $u$ of the moving medium is taken into
account, or the case $u=v$ is considered, i.e., it is supposed that the
observer frame is comoving with medium, or both decompositions (\ref{E2}),
i.e., (\ref{fm}), and (\ref{M1}), i.e., (\ref{mp}), are made with the same
velocity vector, either $u$ or $v$.

Such assumptions enable the introduction of the electromagnetic excitation
bivector $\mathcal{H}$, Eq. (\ref{F5}), and, by using (\ref{E2}) and (\ref%
{M1}), one finds the decomposition of $\mathcal{H}$ into the electric and
magnetic excitations (other names of which are \textquotedblleft electric
displacement\textquotedblright\ and \textquotedblleft magnetic field
intensity\textquotedblright )%
\begin{equation}
\mathcal{H=}D\wedge u/c+(IH)\cdot u/c^{2},  \label{h1}
\end{equation}%
where, as usual, the electric displacement vector $D=\varepsilon _{0}E+P$
and the magnetic field intensity vector $H=(1/\mu _{0})B-M$ are introduced.
The bivector $\mathcal{H}$ in (\ref{h1}) can be written as a CBGQ in the $%
\{\gamma _{\mu }\}$ basis

\begin{equation*}
\mathcal{H}=(1/2)\mathcal{H}^{\mu \nu }\gamma _{\mu }\wedge \gamma _{\nu },\
\mathcal{H}^{\mu \nu }=(1/c)(D^{\mu }u^{\nu }-D^{\nu }u^{\mu
})+(1/c^{2})\varepsilon ^{\mu \nu \alpha \beta }u_{\alpha }H_{\beta }.
\end{equation*}%
The decomposition (\ref{h1}) was first introduced by Minkowski, Eq. (56)
Sec. 11.6, [1]. Notice that Minkowski dealt only with bulk velocity vector
of the medium $u$; in both his Eqs. (55) (our Eq. (\ref{E2}) but with $v=u$)
and (56) (our Eq. (\ref{h1})) the vector $w$ (our $u$) appears. The same
treatment with the decomposition of $\mathcal{H}$ and consequently with only
one velocity, the velocity $u$, is used in several textbooks, e.g., [12],
and papers, e.g., [13, 14]. However, in general, $u\neq v$, e.g., the
observers are at rest in the laboratory frame ($v=c\gamma _{0}$) in which
the considered medium is moving with velocity $u$ ($u\neq c\gamma _{0}$).
Therefore, we continue with an alternative approach which deals with two
different velocity vectors $v$ and $u$, i.e., with Eqs. (\ref{E2}) and (\ref%
{M1}).

Inserting (\ref{E2}) and (\ref{M1}) into the field equation (\ref{F4}) one
gets the general form of the field equation for a magnetized and polarized
moving medium expressed in terms of $E(x)$, $B(x)$, $P(x)$ and $M(x)$
\begin{equation}
\partial \{\varepsilon _{0}[E\wedge v/c+(IB)\cdot v]+[P\wedge
u/c+(1/c^{2})(MI)\cdot u]\}=j^{(C)}/c.  \label{F7}
\end{equation}%
In the same way as in (\ref{F4}), Eq. (\ref{F7}) with the geometric product
can be divided into the vector part (with sources)
\begin{equation}
\partial \cdot \{\varepsilon _{0}[E\wedge v/c+(IB)\cdot v]+[P\wedge
u/c+(1/c^{2})(MI)\cdot u]\}=j^{(C)}/c  \label{A1}
\end{equation}%
and the trivector part (without sources)%
\begin{equation}
\partial \wedge \lbrack E\wedge v/c+(IB)\cdot v]=0.  \label{A2}
\end{equation}%
The field equation without sources (\ref{A2}) remains unchanged relative to
the corresponding equation for vacuum, because the same assertion holds for
the trivector part of (\ref{F4}). We call (\ref{F7}), i.e., (\ref{A1}) and (%
\ref{A2}), the field equations in the Amp\`{e}rian form, in analogy with
Maxwell's equations when they are written in terms of the 3D vectors $%
\mathbf{E}$, $\mathbf{B}$, $\mathbf{P}$ and $\mathbf{M}$; for the latter
ones and the name see, e.g., Eqs. (4.5) in [15]. Observe that in (\ref{F7}),
i.e., (\ref{A1}), there are two different velocities $u$ and $v$. The
equation (\ref{F7}) is a fundamental result, which is not previously
reported in the physics literature, as I am aware.

If the geometric product is used then there is \emph{only one equation}
\emph{for electromagnetism in moving media, Eq.} (\ref{F4}), i.e., in the Amp%
\`{e}rian form Eq. (\ref{F7}). They are written with abstract 4D geometric
quantities and \emph{they comprise and generalize all usual Maxwell's
equations (with 3D vectors) for moving media.}

If Eq. (\ref{A1}) is written in the $\left\{ \gamma _{\mu }\right\} $ basis,
it becomes
\begin{equation}
\partial _{\alpha }\{\varepsilon _{0}[\delta _{\quad \mu \nu }^{\alpha \beta
}E^{\mu }v^{\nu }+c\varepsilon ^{\alpha \beta \mu \nu }v_{\mu }B_{\nu
}]+[\delta _{\quad \mu \nu }^{\alpha \beta }P^{\mu }u^{\nu
}+(1/c)\varepsilon ^{\alpha \beta \mu \nu }M_{\mu }u_{\nu }]\}\gamma _{\beta
}=j^{(C)\beta }\gamma _{\beta },  \label{I1}
\end{equation}%
where $\delta _{\quad \mu \nu }^{\alpha \beta }=\delta _{\,\,\mu }^{\alpha
}\delta _{\,\,\nu }^{\beta }-\delta _{\,\,\nu }^{\alpha }\delta _{\,\,\mu
}^{\beta }$. Similarly, in the $\left\{ \gamma _{\mu }\right\} $ basis, (\ref%
{A2}) becomes
\begin{equation}
\partial _{\alpha }(c\delta _{\quad \mu \nu }^{\alpha \beta }B^{\mu }v^{\nu
}+\varepsilon ^{\alpha \beta \mu \nu }E_{\mu }v_{\nu })\gamma _{5}\gamma
_{\beta }=0.  \label{I2}
\end{equation}%
Again, as for (\ref{A2}), Eq. (\ref{I2}) is the same as in vacuum. In (\ref%
{I1}), as in (\ref{F7}), i.e., (\ref{A1}), there are two different
velocities $u$ and $v$. The equation (\ref{I1}) does not appear in the
entire previous literature. The equation (\ref{F7}), i.e., (\ref{A1}) and (%
\ref{A2}) and also (\ref{I1}) and (\ref{I2}) are the fundamental results
that are obtained in this paper and they enable an alternative, but viable,
treatment of electromagnetism of moving media.

The equation (\ref{A1}) can be written in another form, i.e., in the
\textquotedblleft source representation\textquotedblright\ as with $F$ and $%
\mathcal{M}$, Eq. (\ref{Fj}),%
\begin{equation}
\partial \cdot \{\varepsilon _{0}[E\wedge v/c+(IB)\cdot
v]\}=j^{(C)}/c-\partial \cdot \lbrack P\wedge u/c+(1/c^{2})(MI)\cdot u],
\label{Ej}
\end{equation}%
according to which the sources of $E$ and $B$ fields are the true current
density $j^{(C)}$ and the $P$ and $M$ vectors.

If the abstract quantities, e.g., $j^{(C)}$, $\partial $, $E$ , etc. in (\ref%
{Ej}) are replaced by their representations, i.e., CBGQs in the standard
basis, $j^{(C)}=j^{(C)\beta }\gamma _{\beta }$, $\partial =\gamma ^{\beta
}\partial _{\beta }$, $E=E^{\mu }\gamma _{\mu }$, ..., then (\ref{Ej})
becomes%
\begin{eqnarray}
&&\partial _{\alpha }\{\varepsilon _{0}[\delta _{\quad \mu \nu }^{\alpha
\beta }E^{\mu }v^{\nu }+c\varepsilon ^{\alpha \beta \mu \nu }v_{\mu }B_{\nu
}]\}\gamma _{\beta }  \notag \\
&=&\{j^{(C)\beta }-\partial _{\alpha }[\delta _{\quad \mu \nu }^{\alpha
\beta }P^{\mu }u^{\nu }+(1/c)\varepsilon ^{\alpha \beta \mu \nu }M_{\mu
}u_{\nu }]\}\gamma _{\beta }.  \label{J}
\end{eqnarray}%
From the equations (\ref{A1}) or (\ref{Ej}),\textit{\ }i.e., with CBGQs, (%
\ref{I1}) or (\ref{J}), it is clear that \emph{in 4D spacetime it is not
possible to separate the field equation with sources for the} $E$ \emph{%
field from that one for the }$B$ \emph{field.} Thus, \emph{the usual Amp\`{e}%
re-Maxwell law and Gauss's law are inseparably connected in one law} - \emph{%
Eq. }(\ref{A1}) or \emph{Eq. }(\ref{Ej}), i.e., (\ref{I1}) or (\ref{J}).
Similarly, \emph{in }(\ref{A2}), i.e., Eq.\emph{\ }(\ref{I2}), \emph{%
Faraday's law and the law that expresses the absence of magnetic charge are
also inseparably connected in one law.} This is an essential difference
relative to Maxwell's equations with 3D vectors $\mathbf{E}$, $\mathbf{B}$, $%
\mathbf{P}$ and $\mathbf{M}$. Of course, the same statement holds for the
original Eq. (\ref{R1}), i.e., for the vacuum as well.

The mathematical reason for such an inseparability is that, e.g., the
gradient operator $\partial $ is a vector field defined on 4D spacetime. If
represented in some basis then its vector character remains unchanged only
when \emph{all its components together with associated basis vectors} are
taken into account in the considered equation. The same holds for other
vectors $E$, $B$, $j$, $P$, etc. and multivectors like $F$, $\mathcal{M}$,
... . For example, in general, in 4D spacetime, the current density vector $%
j $ is a well-defined physical quantity, but it is not the case with the
usual charge density $\rho $ and the usual current density $\mathbf{j}$ as a
3-vector. Similarly, in general, the gradient operator $\partial $ cannot be
divided into the usual time derivation and the spatial derivations. In 4D
spacetime, an independent physical reality is attributed to the position
vector $x$, the gradient operator $\partial $, the current density vector $j$%
, the vectors of the electric and magnetic fields $E$ and $B$, respectively,
etc., but not to the 3-vector $\mathbf{r}$ and the time $t$, to 3D vectors $%
\mathbf{j}$, $\mathbf{E}$, $\mathbf{B}$, etc. Therefore, in 4D spacetime, it
is not possible to speak about the static case in electromagnetism, i.e.,
about the electrostatics and magnetostatics.

An important consequence stems from the above mentioned inseparability of 4D
spacetime into the 3D space and the time and therefore from the
inseparability of Eq. (\ref{A1}), i.e. (\ref{Ej}), into two laws, and
similarly for Eq. (\ref{A2}). It can be seen from Maxwell's equations with
3D vectors, e.g., Eqs. (\ref{p4}) and (\ref{p5}), and also (\ref{g2}) and (%
\ref{g3}), which all are given below, that \emph{in the static case} the
electric and magnetic fields, $\mathbf{E}$ and $\mathbf{B}$, respectively,
are completely decoupled. However, as already stated, in 4D spacetime there
is no static case. The equations (\ref{A1}), i.e., (\ref{Ej}), and (\ref{A2}%
) reveal that \emph{the vectors of the electric and magnetic fields} $E$
\emph{and} $B$, respectively, \emph{are never decoupled.} This statement
holds for the vacuum as well. Thus if, for example, we have a magnetization $%
M$ (a permanent magnet) but with a negligible permanent polarization $P$ and
without $j^{(C)}$, then, as can be seen from (\ref{Ej}), $M$ will induce
\emph{both} $B$ \emph{and} $E$. Such a result is completely understandable
because $E$ and $B$ are derived from \emph{one} fundamental quantity, the
electromagnetic field bivector $F$, by the decomposition of $F$ (\ref{E2})
and by (\ref{E1}), and similarly $P$ and $M$ are derived from \emph{one}
quantity, the generalized magnetization-polarization bivector $\mathcal{M}$,
by the decomposition of $\mathcal{M}$ (\ref{M1}) and by (\ref{M2}). The
equations (\ref{F4}), i.e., their \textquotedblleft source
representation\textquotedblright\ (\ref{Fj}), are the basic field equations
with the bivectors $F$ and $\mathcal{M}$; $F$ unites $E$ and $B$ and $%
\mathcal{M}$ unites $P$ and $M$. Besides, $F$ is independent on $v$ and $%
\mathcal{M}$ is independent on $u$. The formulation of electromagnetism of
moving media could be done exclusively in terms of $F$ and $\mathcal{M}$ in
the same way as in [2] for vacuum.

It is worth mentioning that in the integral form the equation that
corresponds to the local equation (\ref{A1}) can be obtained from Eq. (21)
in [2] replacing $F$ by $F+\mathcal{M}/\varepsilon _{0}$, $j$ by $j^{(C)}$
and inserting into it the decompositions (\ref{E2}) and (\ref{M1}), and
similarly for (\ref{A2}) and Eq. (18) in [2].\bigskip \bigskip

\noindent \textbf{4. Observers are at rest in a stationary medium. Comparison%
}

\textbf{with the usual formulation with 3D vectors}\bigskip \bigskip

\noindent Recently, [16 - 21], [11], it is proved that, contrary to the
general belief, the UT of the 3D $\mathbf{E(r,}t\mathbf{)}$ and $\mathbf{B(r,%
}t\mathbf{)}$, see, e.g., Eqs. (11.148) and (11.149) in [6], or Eq. (6) in
[5], i.e., Eq. (\ref{JCB}) here, \emph{differ} from the LT (boosts) of the
4D $E$ and $B$ vector fields. Note that, in previous literature, starting
with Einstein [3], the UT of $\mathbf{E}$ and $\mathbf{B}$ are always
considered to be the relativistically correct LT. The same holds for the UT
of $\mathbf{P}$ and $\mathbf{M}$, Eq. (\ref{ps}) below, and the LT of
vectors $P$ and $M$. For a recent more detailed review see Secs. 3.1 and 3.2
in [22]. The essential point is that in the UT \emph{the transformed} $%
\mathbf{E}^{\prime }$ \emph{is expressed by the mixture of the 3D vectors} $%
\mathbf{E}$ \emph{and} $\mathbf{B}$, Eq. (11.149) in [6], which is
\begin{equation}
\mathbf{E}^{\prime }=\gamma (\mathbf{E}+\mathbf{\beta \times }c\mathbf{B)-}%
(\gamma ^{2}/(1+\gamma ))\mathbf{\beta (\beta \cdot E),}  \label{JCB}
\end{equation}%
\emph{and similarly for} $\mathbf{B}^{\prime }$. For the components
implicitly taken in the standard basis it holds that
\begin{equation}
E_{1}^{\prime }=E_{1},\quad E_{2}^{\prime }=\gamma (E_{2}-\beta
cB_{3}),\quad E_{3}^{\prime }=\gamma (E_{3}+\beta cB_{2}),  \label{je}
\end{equation}%
and similarly for the components of $\mathbf{B}^{\prime }$, Eq. (11.148) in
[6].

The same holds for the couple of the 3D vectors $\mathbf{P}$ and $\mathbf{M}$%
\ and their UT

\begin{eqnarray}
\mathbf{P} &=&\gamma (\mathbf{P}^{\prime }+\mathbf{\beta \times M}^{\prime
}/c\mathbf{)-}(\gamma ^{2}/(1+\gamma ))\mathbf{\beta (\beta \cdot P}^{\prime
}\mathbf{),}  \notag \\
\mathbf{M} &=&\gamma (\mathbf{M}^{\prime }-\mathbf{\beta \times }c\mathbf{P}%
^{\prime }\mathbf{)-}(\gamma ^{2}/(1+\gamma ))\mathbf{\beta (\beta \cdot M}%
^{\prime }\mathbf{).}  \label{ps}
\end{eqnarray}%
These UT of $\mathbf{P}$ and $\mathbf{M}$, Eq. (\ref{ps}), are given, e.g.,
by Eq. (4.2) in [15].

On the other hand, in ISR, as shown in [16 - 21], [11], \emph{the correct LT
always transform the 4D algebraic object representing, e.g., the electric
field only to the electric field; there is no mixing with the magnetic
field. }These correct LT are given by, e.g., Eq. (8) in [5], i.e., Eq. (13)
in [11]. The same happens with $P$ and $M$. The LT of the components $E^{\mu
}$ (in the $\{\gamma _{\mu }\}$ basis) of $E=E^{\mu }\gamma _{\mu }$ are
given as%
\begin{equation}
E^{\prime 0}=\gamma (E^{0}-\beta E^{1}),\ E^{\prime 1}=\gamma (E^{1}-\beta
E^{0}),\ E^{\prime 2,3}=E^{2,3},  \label{ELT}
\end{equation}%
for a boost along the $x^{1}$ axis. As already mentioned, any CBGQ is
unchanged under the LT, i.e., it holds that $E=E^{\nu }\gamma _{\nu
}=E^{\prime \nu }\gamma _{\nu }^{\prime }=E_{r}^{\nu }r_{\nu }=E_{r}^{\prime
\nu }r_{\nu }^{\prime }$, where the primed quantities in both bases $%
\{\gamma _{\mu }\}$ and $\{r_{\mu }\}$ are the Lorentz transforms of the
unprimed ones. For the $\{r_{\mu }\}$ basis, with the \textquotedblleft
radio,\textquotedblright\ \textquotedblleft r\textquotedblright
synchronization, and the LT in that basis, see [23] and [24]. The same LT
hold for any other vector, e.g., $x$, $B$, $P$, $M$, etc.

A short derivation of the above LT is presented in [21]. Let us introduce
the $\gamma _{0}$-frame in which the standard basis is chosen and $v=c\gamma
_{0}$. Therefore, in the $\gamma _{0}$-frame, $E$ from (\ref{E1}) becomes $%
E=F\cdot \gamma _{0}$. If, in the $\gamma _{0}$-frame, that $E$ is written
as a CBGQ, then the components are given as $E^{0}=B^{0}=0$ and only the
spatial components remain, $E^{i}=F^{i0}$, $B^{i}=(1/2c)\varepsilon
^{ijk0}F_{jk}$; the same components as in, e.g., Eq. (11.137) in [6]. It is
proved by Minkowski [1], and reinvented and generalized in [16 - 21], [11],
see also Sec. 5 in [22], that in the mathematically correct procedure for
the derivation of the LT of $E$ \emph{both} $F$ \emph{and} the velocity
vector $v$ have to be transformed by the LT, e.g., for the LT from the $%
\gamma _{0}$-frame;%
\begin{equation}
E=E^{\mu }\gamma _{\mu }=[(1/c)F^{i0}v_{0}]\gamma _{i}=[(1/c)F^{\prime \mu
\nu }v_{\nu }^{\prime }]\gamma _{\mu }^{\prime }=E^{\prime \mu }\gamma _{\mu
}^{\prime }.  \label{elt}
\end{equation}%
The velocity vector $v$ transforms as any other vector and it holds that $%
v=v^{0}\gamma _{0}=c\gamma _{0}=v^{\prime \mu }\gamma _{\mu }^{\prime }$.
Hence, \emph{the components} $E^{\mu }$ \emph{transform by the LT again to
the components} $E^{\prime \mu }$ \emph{of the same electric field vector},
i.e., the above quoted LT (\ref{ELT}) of the components $E^{\prime \mu }$
are obtained. \emph{The main point is that the transformed components} $%
E^{\prime \mu }$ \emph{are not determined only by} $F^{\prime \mu \nu }$,%
\emph{\ as in all usual approaches}, e.g., Eqs. (11.147) and (11.148) in
[6], \emph{but also by} $v^{\prime \mu }$.

As already stated, Minkowski, section 11.6 in [1], was the first who derived
these mathematically correct LT. He assumed that $v$, $E$ and $B$ are $%
1\times 4$ matrices and $F$ is a $4\times 4$ matrix; their components are
implicitly determined in the standard basis. He described how $v$ and $F$
separately transform under the LT $A$ (the matrix of the LT is denoted as $A$%
)

\begin{equation}
v\longrightarrow v^{\prime }=vA,\quad F\longrightarrow F^{\prime }=A^{-1}FA.
\label{vF}
\end{equation}%
Then, as shown by Minkowski, the mathematically correct LT of $E=vF$ is
\begin{equation}
E=vF\longrightarrow E^{\prime }=(vA)(A^{-1}FA)=(vF)A=EA.  \label{LE}
\end{equation}%
Thus, under the LT \emph{both} quantities, the field-strength tensor $F$\ ($%
4\times 4$ matrix) \emph{and the 4-velocity} $v$ ($1\times 4$ matrix)\ are
transformed and their product transforms as any $1\times 4$ matrix
transforms. It is already mentioned that this mathematically correct
procedure is reinvented and generalized using 4D geometric quantities both
in the tensor formalism and in the geometric algebra formalism in [16 - 21],
[11].

The comparison with experiments in electromagnetism, the motional emf [17]
and Secs. 7 - 7.2 here, the Faraday disk [18], and the Trouton-Noble
experiment [2, 25], shows that the approach with 4D geometric quantities and
their LT, i.e., the ISR, always agrees with the principle of relativity and
it is in a true agreement (independent of the chosen inertial reference
frame and of the chosen system of coordinates in it) with all experiments in
electromagnetism. Also, it is shown in the mentioned papers that such a true
agreement does not exist in the usual formulations of SR, e.g., [6, 15, 26],
in which the electric and magnetic fields are represented by the 3D vectors $%
\mathbf{E(r,}t\mathbf{)}$ and $\mathbf{B(r,}t\mathbf{)}$ that transform
according to the UT. The same conclusion about the true agreement between
the approach with 4D geometric quantities, the ISR, and the well-known
experiments that test special relativity is obtained in [27]. There, in
[27], and particularly in [23], it is explicitly shown that the relativity
of simultaneity, the Lorentz contraction and the time dilation are not
well-defined in 4D spacetime. They are not the intrinsic relativistic
effects, because they depend on the chosen synchronization. But, \emph{every
synchronization is only a convention and physics must not depend on
conventions. This true agreement of ISR with experiments directly proves the
physical reality of 4D geometric quantities.}

Having briefly discussed the LT and the UT we go back to the discussion of
Eqs. (\ref{I1}) (i.e., (\ref{J})) and (\ref{I2}). In all relatively moving
inertial frames of reference and for any system of coordinates in them every
term in the considered equations is always the same, because all CBGQs are
the Lorentz invariant quantities, e.g., $j^{(C)\beta }\gamma _{\beta
}=j^{\prime (C)\beta }\gamma _{\beta }^{\prime }=j_{r}^{(C)\beta }\gamma
_{r,\beta }=$ ... . Observe that, in 4D spacetime, \emph{only if all
components, together with the associated basis vectors,} are taken into
account in every term then all terms are invariant under the passive LT and
thus Eqs. (\ref{I1}) ((\ref{J})) and (\ref{I2}) remain unchanged for
different relatively moving inertial frames and for different systems of
coordinates in them. Only in that case the physical quantities and the
equations with them are correctly defined in 4D spacetime and the principle
of relativity is naturally satisfied. This means that, in general, it is not
allowed to consider separately some parts of 4D geometric quantities, or
some parts of the equations with them, e.g., to take the part with $\gamma
_{0}$ separately from those ones with $\gamma _{i}$ in Eqs. (\ref{I1}) ((\ref%
{J})) and (\ref{I2}). Thus, for example, in 4D spacetime, only the whole
current density $j^{(C)}$, the abstract vector from (\ref{A1})\emph{\ }and (%
\ref{Ej}), or some of its representations, e.g., that one in the standard
basis, the CBGQ, $j=j^{(C)\beta }\gamma _{\beta }$, is well-defined physical
quantity, but not the charge density, $j^{(C)0}$ component, or the spatial
components $j^{(C)i}$ taken alone. From the viewpoint of this 4D geometric
approach the physical meaning of the charge density $\rho $ is not
well-defined. It is the temporal component $j^{0}/c$ for one observer, but
it transforms by the LT to the temporal component \emph{and} the spatial
component as well for the relatively moving observer. The same holds for the
gradient operator $\partial $ and its CBGQ in the standard basis $\partial
=\gamma ^{\beta }\partial _{\beta }$ and for all other 4D geometric
quantities.

This is particularly visible going to some nonstandard basis, like the $%
\{r_{\mu }\}$ basis, i.e. with the \textquotedblleft
radio,\textquotedblright\ \textquotedblleft r,\textquotedblright\
synchronization, see, e.g., [23], Preprints in [27] and [24]. The
\textquotedblleft r\textquotedblright\ synchronization is commonly used in
everyday life. If the observers who are at different distances from the
studio clock set their clocks by the announcement from the studio\ then they
have synchronized their clocks with the studio clock according to the
\textquotedblleft r\textquotedblright\ synchronization. The unit vectors in
the $\{\gamma _{\mu }\}$ basis and the $\{r_{\mu }\}$ basis are connected as
$r_{0}=\gamma _{0}$, $r_{i}=\gamma _{0}+\gamma _{i}$. The components of any
vector are connected in the same way as the components of the position
vector $x$ are connected, $x_{r}^{0}=x^{0}-x^{1}-x^{2}-x^{3}$, $%
x_{r}^{i}=x^{i}$, e.g. for the components of vector $E$ it also holds that $%
E_{r}^{0}=E^{0}-E^{1}-E^{2}-E^{3}$, $E_{r}^{i}=E^{i}$.\ The inverse
relations are $\gamma _{0}=r_{0}$, $\gamma _{i}=r_{i}-r_{0}$ and, e.g., for
the components of the current density vector $j$, $%
j^{0}=j_{r}^{0}+j_{r}^{1}+j_{r}^{2}+j_{r}^{3}$, $j^{i}=j_{r}^{i}$. Thus,
even in the same frame, the charge density in the $\{\gamma _{\mu }\}$ basis
($j^{0}=c\rho $) loses its usual meaning; it is expressed by the sum of all
components in the $\{r_{\mu }\}$ basis. However, observe that, as already
stated, $j=j^{\mu }\gamma _{\mu }=j_{r}^{\mu }r_{\mu }$ and the same holds
for $E=E^{\mu }\gamma _{\mu }=E_{r}^{\mu }r_{\mu }$, for $B$, for $x$, etc.
This reveals that in the $\{r_{\mu }\}$ basis the space and time cannot be
separated. Hence, in 4D spacetime the usual interpretations of the physical
quantities, e.g., the charge density $\rho $ and the current density as a
3-vector $\mathbf{j}$, are inappropriate.

\emph{An independent physical reality can be attributed either to the
abstract geometric quantities, }e.g.,\emph{\ vectors} $x$, $E$, $B$, $P$, $M$%
, $j$, .. \emph{bivectors} $F$, $\mathcal{M}$, .., \emph{or to their
representations in different bases, the CBGQs, like} $j^{\mu }\gamma _{\mu }$%
, $E_{r}^{\mu }r_{\mu }$, $M^{\prime \beta }\gamma _{\beta }^{\prime }$, etc.

Now, let us consider Eq. (\ref{I1}) in the case when $u=v$, i.e., the
observer frame is comoving with medium. In that case it can be taken that $%
v=u=c\gamma _{0}$ ($u^{\mu }=v^{\mu }=(c,0,0,0)$), i.e., that the observers
who measure fields are at rest in a stationary medium. Then, Eq. (\ref{I1})
becomes
\begin{eqnarray}
\partial _{\alpha }\{\varepsilon _{0}[\delta _{\quad \mu \nu }^{\alpha \beta
}E^{\mu }(\gamma _{0})^{\nu }+c\varepsilon ^{\alpha \beta \mu \nu }(\gamma
_{0})_{\mu }B_{\nu }] &&+  \notag \\
\lbrack \delta _{\quad \mu \nu }^{\alpha \beta }P^{\mu }(\gamma _{0})^{\nu
}+(1/c)\varepsilon ^{\alpha \beta \mu \nu }M_{\mu }(\gamma _{0})_{\nu
}]\}\gamma _{\beta } &=&c^{-1}j^{(C)\beta }\gamma _{\beta },  \label{I3}
\end{eqnarray}%
The equation (\ref{I3}) can be also written in the \textquotedblleft source
representation\textquotedblright\ as
\begin{eqnarray}
&&\partial _{\alpha }\{\varepsilon _{0}[\delta _{\quad \mu \nu }^{\alpha
\beta }E^{\mu }(\gamma _{0})^{\nu }+c\varepsilon ^{\alpha \beta \mu \nu
}(\gamma _{0})_{\mu }B_{\nu }]\}\gamma _{\beta }  \notag \\
&=&\{c^{-1}j^{(C)\beta }-\partial _{\alpha }[\delta _{\quad \mu \nu
}^{\alpha \beta }P^{\mu }(\gamma _{0})^{\nu }+(1/c)\varepsilon ^{\alpha
\beta \mu \nu }M_{\mu }(\gamma _{0})_{\nu }]\}\gamma _{\beta }.  \label{i4}
\end{eqnarray}%
As already mentioned, the sources of both fields together, $E$ and $B$, are
the true current density $j^{(C)}$ and the polarization and magnetization
vectors, $P$ and $M$ respectively.

The observer frame is the $\gamma _{0}$-frame, $v=c\gamma _{0}$, which, with
(\ref{E1}), yields that $E^{0}=B^{0}=0$ and $E^{i}=F^{i0}$, $%
B^{i}=(1/2c)\varepsilon ^{ijk0}F_{jk}$. Furthermore, in the considered case,
the $\gamma _{0}$-frame coincides with the rest frame of the medium. Hence,
in that frame and with (\ref{M2}), it also holds that $P^{0}=M^{0}=0$, $%
P^{i}=\mathcal{M}^{i0}$, $M^{i}=(c/2)\varepsilon ^{0ijk}\mathcal{M}_{jk}$.
Then, Eq. (\ref{I3}) becomes
\begin{eqnarray}
\lbrack \partial _{k}(E^{k}+P^{k}/\varepsilon _{0})-j^{(C)0}/c\varepsilon
_{0}]\gamma _{0}+ &&\{c\varepsilon ^{ijk0}\partial _{j}[(B_{k}-\mu _{0}M_{k}]
\notag \\
-j^{(C)i}/c\varepsilon _{0}-\partial _{0}(E^{i}+P^{i}/\varepsilon
_{0})\}\gamma _{i} &=&0  \label{e}
\end{eqnarray}%
In vacuum, Eq. (\ref{e}) coincides with the first two terms, i.e., the terms
with $\gamma _{0}$ and $\gamma _{i}$, in Eq. (8) in [18]. In the approach
with 4D geometric quantities, i.e., in the ISR, it is not possible to make
any further simplification. \emph{In 4D spacetime}, \emph{only the whole Eq.}
(\ref{e}) \emph{is physically meaningful and there is no physical sense in
some parts of it, for example, to take the part with} $\gamma _{0}$ \emph{%
separately from those ones with} $\gamma _{i}$. Note that in the approach
with 3D vectors there is not any Maxwell's equation that corresponds to Eq. (%
\ref{e}).

Let us write Eq. (\ref{e}) as $a^{\alpha }\gamma _{\alpha }=0$, in which, as
can be easily recognized, the coefficients $a^{\alpha }$ correspond to the
usual Maxwell's equations in the component form. There are \emph{two} \emph{%
Maxwell equations} in the \emph{component form}; the coefficient $a^{0}$
corresponds to the \emph{component form }of the Gauss law for the electric
field and the coefficients $a^{i}$ correspond to the Amp\`{e}re-Maxwell law
in the \emph{component form}. In [18], for the first time, a fundamental
discovery is achieved that the usual Maxwell's equations with 3D vectors
(for vacuum) are not covariant under the LT. In Sec. 2.3 in [18], the active
LT (Eq. (16) in [18]) are applied to Eq. (8) in [18] (in vacuum, as already
stated, our Eq. (\ref{e}) corresponds to the first two terms of Eq. (8) from
[18]). There, in that section, it is obtained that the coefficient $a^{0}$,
which corresponds to the \emph{component form }of the Gauss law for the
electric field, does not transform by the LT again to the Gauss law but to $%
a^{\prime 0}$, $a^{\prime 0}=\gamma a^{0}-\beta \gamma a^{1}$, which is a
combination of the Gauss law and a part of the Amp\`{e}re-Maxwell law ($%
a^{1} $). (In our case, for the material medium, $a^{0}=\partial
_{k}(E^{k}+P^{k}/\varepsilon _{0})-j^{(C)0}/c\varepsilon _{0}$.) If the
Lorentz transformed Eq. (\ref{e}), similarly as in Eqs. (21)-(24) in [18],
is expressed in terms of Lorentz transformed derivatives and Lorentz
transformed vectors $E$, $B$, $P$, $M$ and $j$, then we find the same
equation for $a^{\prime 0}$ as it is Eq. (24) in [18], $a^{\prime
0}=\{[\gamma (\partial _{k}^{\prime }E^{\prime k})-j^{\prime 0}/c\varepsilon
_{0}]+\beta \gamma \lbrack \partial _{1}^{\prime }E^{\prime 0}+c(\partial
_{2}^{\prime }B_{3}^{\prime }-\partial _{3}^{\prime }B_{2}^{\prime })]\}$,
but $E^{\prime \alpha }$ has to be replaced by $E^{\prime \alpha }+P^{\prime
\alpha }/\varepsilon _{0}$, $B_{k}^{\prime }$ by $B_{k}^{\prime }-\mu
_{0}M_{k}^{\prime }$, and $j^{\prime 0}$ by $j^{\prime (C)0}$. The same
discussion holds here as it is presented after Eq. (24) in [18]. From that
discussion, and the above mentioned replacements, one concludes that the LT
do not transform the Gauss law into the \textquotedblleft
primed\textquotedblright\ Gauss law but into quite different law; $a^{\prime
0}$ contains the time component $E^{\prime 0}+P^{\prime 0}/\varepsilon _{0}$%
, whereas $E^{0}=P^{0}=0$, and also the new \textquotedblleft Gauss
law\textquotedblright\ includes the derivatives of the magnetic field. The
same situation happens with other Lorentz transformed terms, which again
explicitly shows that \emph{the Lorentz transformed Maxwell's equations are
not of the same form as the original ones}. Hence, contrary to all previous
considerations, and contrary to the general opinion, \emph{the usual
Maxwell's equations are not Lorentz covariant equations either in vacuum or
in a material medium.} This result proves in another way that \emph{in 4D
spacetime} only the whole Eq. (\ref{e}) is physically meaningful and not its
separate parts. Remember that Eq. (\ref{e}) is derived from Eq. (\ref{I3}),
i.e. from (\ref{I1}), for which it also holds that $a^{\beta }\gamma _{\beta
}=a^{\prime \beta }\gamma _{\beta }^{\prime }=a_{r}^{\beta }\gamma _{r,\beta
}=$ ... . Here, as before, the primed quantities are the Lorentz transforms
of the unprimed ones in the $\{\gamma _{\mu }\}$ basis, whereas the last
expression refers to the $\{r_{\mu }\}$ basis with the \textquotedblleft
r\textquotedblright\ synchronization.

Let us see how Eq. (\ref{e}) could be compared with the usual form of
Maxwell's equations for stationary media, which deals with 3D vectors, e.g.,
[15, 26, 28, 29], etc. Obviously, the comparison will be possible \emph{only}
if the term with $\gamma _{0}$ is considered separately from those ones with
$\gamma _{i}$ and if in Eq. (\ref{e}) \emph{only the components are taken
into account. }But, as explained above, \emph{from the viewpoint of the
geometric approach, }i.e.\emph{, the ISR, such a procedure is not correct in}
\emph{4D spacetime}.

If in Eq. (\ref{e}), i.e., in $a^{0}\gamma _{0}+a^{i}\gamma _{i}=0$, one
takes that $a^{0}=0$, and multiply the spatial components of $E$, $P$ and $%
j^{(C)}$ from $a^{0}$ by \emph{the unit 3D vectors} $\mathbf{i}$, $\mathbf{j}
$, $\mathbf{k}$, then the term with $\gamma _{0}$ will become the equation
\begin{equation}
\nabla \cdot \varepsilon _{0}\mathbf{E}(\mathbf{r},t)=\rho ^{(C)}(\mathbf{r}%
,t)-\nabla \cdot \mathbf{P}(\mathbf{r},t),  \label{p4}
\end{equation}%
what is Eq. (4.5) (3) in [15], or Eq. (9-6)(1) in [28], or Eq. (4.139) in
[29], etc. In the same way it will be obtained that the terms with $\gamma
_{i}$ will become the equation
\begin{equation}
\nabla \times \mathbf{B}(\mathbf{r},t)=\mu _{0}[\mathbf{j}^{(C)}(\mathbf{r}%
,t)+\partial (\varepsilon _{0}\mathbf{E}(\mathbf{r},t)+\mathbf{P(\mathbf{r},}%
t\mathbf{))}/\partial t+\nabla \times \mathbf{M}(\mathbf{r},t)],  \label{p5}
\end{equation}%
what is Eq. (4.5) (2) in [15], or Eq. (9-6)(4) in [28], or Eq. (4.142) in
[29], etc. The way in which the equations with 3D vectors (\ref{p4}) and (%
\ref{p5}) are constructed clearly shows that Eq. (\ref{e}) is essentially
different than Eqs. (\ref{p4}) and (\ref{p5}).

Similarly, we find that in the $\gamma _{0}$-frame Eq. (\ref{I2}) becomes
\begin{equation}
(c^{2}\partial _{k}B^{k})\gamma _{5}\gamma _{0}-(c\partial
_{0}B^{i}+\varepsilon ^{ijk0}\partial _{j}E_{k})\gamma _{5}\gamma _{i}=0.
\label{gi}
\end{equation}%
The equation (\ref{gi}) coincides, without any changes, with the last two
terms, i.e., the terms with $\gamma _{5}\gamma _{0}$ and $\gamma _{5}\gamma
_{i}$, in the equation for vacuum, Eq. (8) in [18]. As already stated, in (%
\ref{gi}), Faraday's law and the law that expresses the absence of magnetic
charge are \emph{inseparably} connected in one law. But, as in the
discussion of Eq. (\ref{e}), we can make the comparison of Eq. (\ref{gi})
with the usual form of Maxwell's equations for stationary media, which deals
with 3D vectors. The equation with 3D vectors that expresses the absence of
magnetic charge%
\begin{equation}
\nabla \cdot \mathbf{B}(\mathbf{r},t)=0,  \label{g2}
\end{equation}%
can be constructed from the term with $\gamma _{5}\gamma _{0}$ in (\ref{gi})
in the same way as Eqs. (\ref{p4}) and (\ref{p5}) are constructed from Eq. (%
\ref{e}). The obtained Eq. (\ref{g2}) is Eq. (4.5) (4) in [15], or Eq.
(9-6)(2) in [28], or Eq. (4.140) in [29], or Eq. (7.55)(ii) in [26], etc.
Similarly, the terms with $\gamma _{5}\gamma _{i}$ will give the equation
with 3D vectors, Faraday's law,%
\begin{equation}
\nabla \times \mathbf{E}(\mathbf{r},t)=-\partial \mathbf{B}(\mathbf{r}%
,t)/\partial t,  \label{g3}
\end{equation}%
what is Eq. (4.5) (1) in [15], or Eq. (9-6)(3) in [28], or Eq. (4.141) in
[29], or Eq. (7.55)(iii) in [26], etc. (The components of vectors $E$, $B$, $%
P$, $M$ with superscripts ($E^{i}$, $B^{i}$, $P^{i}$, $M^{i}$) from (\ref{e}%
) are identified with the components of the usual 3D vectors and $%
\varepsilon ^{0123}=1$.) As already mentioned, Maxwell's equations in terms
of $\mathbf{E}$, $\mathbf{B}$, $\mathbf{P}$ and $\mathbf{M}$, Eqs. (4.5) in
[15], i.e., (\ref{p4}), (\ref{p5}), (\ref{g2}) and (\ref{g3}) here, are said
to be in the Amp\`{e}rian form.

If, as usual, the electric displacement 3D vector $\mathbf{D}=\varepsilon
_{0}\mathbf{E}+\mathbf{P}$ has been introduced together with the magnetic
field intensity 3D vector $\mathbf{H}=(1/\mu _{0})\mathbf{B}-\mathbf{M}$
then Eqs. (\ref{p4}) and (\ref{p5}) become
\begin{eqnarray}
\nabla \cdot \mathbf{D}(\mathbf{r},t) &=&\rho ^{(C)}(\mathbf{r},t),  \notag
\\
\nabla \times \mathbf{H}(\mathbf{r},t) &=&\mathbf{j}^{(C)}(\mathbf{r}%
,t)+\partial \mathbf{D}/\partial t.  \label{dh}
\end{eqnarray}%
In (\ref{dh}), the first equation is Eq. (9-7)(1) in [28], or Eq. (4.139a)
in [29], or Eq. (7.55)(i) in [26], etc., whereas the second one is Eq.
(9-7)(4) in [28], or Eq. (4.142a) in [29], or Eq. (7.55)(iv) in [26], etc.

According to this discussion there is an essential difference between
Maxwell's equations (\ref{p4}), (\ref{p5}), or (\ref{dh}), with the 3D
vectors and Eq. (\ref{e}), i.e., the equations from which (\ref{e}) is
derived, (\ref{I3}), (\ref{I1}) and (\ref{A1}). In the 4D geometric
approach, i.e., in the ISR, there is \emph{one law}, Eq. (\ref{e}), i.e., (%
\ref{I3}), or (\ref{I1}), or (\ref{A1}), whereas in the approach with 3D
vectors there are \emph{two laws}, Eqs. (\ref{p4}) and (\ref{p5}), or (\ref%
{dh}). In order to obtain two laws (\ref{p4}), (\ref{p5}), or (\ref{dh}),
from Eq. (\ref{e}) we had to make several steps. First, the term with $%
\gamma _{0}$ is taken separately from those ones with $\gamma _{i}$, then
only the components in these terms are taken into account and finally the
components are multiplied by the unit 3D vectors $\mathbf{i}$, $\mathbf{j}$,
$\mathbf{k}$. But, in 4D spacetime, as explained above, these steps are not
mathematically correct. This consideration clearly shows that Eq. (\ref{e}),
i.e., Eqs. (\ref{I3}), or (\ref{I1}), or (\ref{A1}), from which (\ref{e}) is
derived, is not equivalent to Eqs. (\ref{p4}), (\ref{p5}), or (\ref{dh}).
The equation (\ref{e}) is more general and, strictly speaking, it is not
possible to obtain Eqs. (\ref{p4}), (\ref{p5}), or (\ref{dh}) from (\ref{e})
by a mathematically correct procedure in 4D spacetime. The same
consideration holds in the same measure for the relation between (\ref{gi})
and the equations with 3D vectors (\ref{g2}) and (\ref{g3}).

Furthermore, in the usual approach with 3D vectors one can speak about the
static case. Then, Eqs. (\ref{p4}) and (\ref{p5}), or the first and the
second equation in (\ref{dh}), are completely decoupled, i.e., in the static
case the electric and magnetic fields as 3D vectors are decoupled. In the 4D
geometric approach such a decoupling is never possible, because there is
only one law in which there are both together $E$ and $B$ as vectors. The
same consideration holds for Maxwell's equations (\ref{g2}) and (\ref{g3})
with 3D vectors and Eq. (\ref{gi}), i.e., the equations from which (\ref{gi}%
) is derived, (\ref{I2}) and (\ref{A2}).

The most important difference is the following. The quantities entering into
(\ref{e}) and (\ref{gi}) are representations in the standard basis of the
abstract 4D quantities from Eq. (\ref{F7}), i.e., Eqs. (\ref{A1}) and (\ref%
{A2}). All these quantities are correctly defined in 4D spacetime and they
correctly transform under the LT, e.g., for the components (\ref{ELT}), or
for the vector $E$, Eq. (8) in [5], i.e., Eq. (13) in [11], whereas it is
not the case with quantities appearing in (\ref{p4}),(\ref{p5}), (\ref{g2}),
(\ref{g3}) and (\ref{dh}), which transform according to the UT, e.g., (\ref%
{JCB}), of the 3D vectors $\mathbf{E}$, $\mathbf{B}$, $\mathbf{D}$, $\mathbf{%
H}$.\bigskip \bigskip

\noindent \noindent \textbf{5. Observers are at rest in the laboratory
frame, but material }

\textbf{medium is moving. Comparison with the usual formulation }

\textbf{with 3-vectors}\bigskip

\noindent Let us examine Eqs. (\ref{I1}) and (\ref{I2}) in the case of a
moving material medium, but the observers are at rest in the laboratory
frame, which will be denoted as the $S$ frame. Then, in $S$, $v=c\gamma _{0}$%
, $v^{\mu }=(c,0,0,0)$). Thus the laboratory frame is the $\gamma _{0}$%
-frame in which it holds that $E^{0}=B^{0}=0$ and $E^{i}=F^{i0}$, $%
B^{i}=(1/2c)\varepsilon ^{ijk0}F_{jk}$. Obviously, Eq. (\ref{I2}) becomes
the same as Eq. (\ref{gi}), i.e., the same as in vacuum and the whole
discussion about the comparison of (\ref{gi}) with Maxwell's equations with
3D vectors remains unchanged. But, it is not so for Eq. (\ref{I1}).

In the $\gamma _{0}$-frame the considered medium is moving with velocity $u$%
, $u\neq c\gamma _{0}$, i.e., some of $u^{i}$ are $\neq 0$. The rest frame
of the medium will be denoted as the $S^{\prime }$ frame. For the sake of
comparison with the usual formulation we present the considered equation in
an expanded form in which the term with $\gamma _{0}$ and the terms with $%
\gamma _{i}$ are explicitly written. Then, in the laboratory frame, which is
the $\gamma _{0}$-frame, Eq. (\ref{I1}) becomes
\begin{eqnarray}
\{\partial _{k}[\varepsilon
_{0}E^{k}+c^{-1}(P^{k}u^{0}-P^{0}u^{k})+c^{-2}\varepsilon
^{kij0}M_{i}u_{j}]-c^{-1}j^{(C)0}\}\gamma _{0} &&+  \notag \\
\{-c^{-1}j^{(C)i}+(c\mu _{0})^{-1}\varepsilon ^{ijk0}\partial
_{j}B_{k}-\varepsilon _{0}\partial _{0}E^{i} &&+  \notag \\
c^{-1}\partial _{\mu }(P^{\mu }u^{i}-P^{i}u^{\mu })-c^{-2}\varepsilon ^{i\mu
\alpha \beta }\partial _{\mu }M_{\alpha }u_{\beta }]\}\gamma _{i} &=&0
\label{j1}
\end{eqnarray}%
Again, as in the discussion of Eq. (\ref{e}), it can be argued that \emph{in
4D spacetime}, \emph{only the whole Eq.} (\ref{j1}) \emph{is physically
meaningful and there is no physical sense in some parts of it, for example,
to take the part with} $\gamma _{0}$ \emph{separately from those ones with} $%
\gamma _{i}$. Observe that in (\ref{j1}) there are terms with $P^{0}$ and $%
M^{0}$, which cannot exist in the usual formulation with 3D vectors.

What will be obtained from (\ref{j1}) for the case of low velocities of the
medium, i.e., for $\beta _{u}\ll 1$, $\gamma _{u}=(1-\beta
_{u}^{2})^{-1/2}\simeq 1$, where, in $S$ and in the $\{\gamma _{\mu }\}$
basis, $u=u^{\nu }\gamma _{\nu }$, $u^{\nu }=(\gamma _{u}c,\gamma
_{u}U^{1},\gamma _{u}U^{2},\gamma _{u}U^{3})$, $U^{k}$ are the same as the
components of the 3-velocity $\mathbf{U}$ and $\beta _{u}=\left\vert \mathbf{%
U}\right\vert /c$. To determine and compare $P^{0}$ and $P^{k}$ in $S$ \emph{%
we use the LT of} $P^{\prime \mu }$ \emph{from} $S^{\prime }$, the rest
frame of the medium, and, for simplicity, it is taken that the medium, the $%
S^{\prime }$ frame, is moving along the common $+x^{1}$, $x^{\prime 1}$
axes, i.e., $u^{\nu }=(\gamma _{u}c,\gamma _{u}U^{1},0,0)$. In $S^{\prime }$%
, $P^{\prime \mu }=(0,P^{\prime 1},P^{\prime 2},P^{\prime 3})$. Then, using
the LT, the same as in Eq. (\ref{ELT}), $P^{\mu }=(\beta _{u}\gamma
_{u}P^{\prime 1},\gamma _{u}P^{\prime 1},P^{\prime 2},P^{\prime 3})$. Since $%
\beta _{u}\ll 1$ and $\gamma _{u}\simeq 1$ it follows that $P^{0}\ll P^{1}$
and $P^{k}u^{0}-P^{0}u^{k}$ in (\ref{j1}) becomes $\simeq cP^{k}$, i.e., in
that approximation $P^{0}u^{k}$ can be neglected relative to $P^{k}u^{0}$.
In the same way it can be concluded that $M^{0}u^{k}$ can be neglected
relative to $M^{k}u^{0}$. Therefore, with these approximations, Eq. (\ref{j1}%
) can be written as
\begin{eqnarray}
\partial _{k}\varepsilon _{0}E^{k}\gamma _{0}+(c\mu _{0})^{-1}\varepsilon
^{ijk0}\partial _{j}B_{k}\gamma _{i}\simeq \lbrack c^{-1}j^{(C)0}-\partial
_{k}P^{k}- &&  \notag \\
c^{-2}\varepsilon ^{kij0}\partial _{k}M_{i}U_{j}]\gamma
_{0}+[c^{-1}j^{(C)i}+\partial _{0}(\varepsilon _{0}E^{i}+P^{i}) &&+  \notag
\\
c^{-1}((U^{k}\partial _{k})P^{i}-U^{i}(\partial
_{k}P^{k}))+c^{-2}\varepsilon ^{ijk0}(\partial _{0}M_{j}U_{k}+c\partial
_{j}M_{k}]\gamma _{i}, &&  \label{j2}
\end{eqnarray}%
Notice that (\ref{j2}) \emph{is obtained from} (\ref{j1}) \emph{using the LT
of the vectors} $P^{\mu }\gamma _{\mu }$ and $M^{\mu }\gamma _{\mu }$ \emph{%
and not the UT of the 3D vectors} $\mathbf{P}$ and $\mathbf{M}$, Eq. (\ref%
{ps}). We see that $D^{k}=\varepsilon _{0}E^{k}+P^{k}$ \emph{and} $%
H^{k}=B^{k}/\mu _{0}-M^{k}$ \emph{can be introduced into Eq.} (\ref{j2}),
\emph{whereas such replacement is not possible for Eq.} (\ref{j1}), \emph{%
due to the existence of the terms with} $P^{0}$ \emph{and} $M^{0}$ \emph{in}
(\ref{j1}). Then the part with $\gamma _{0}$ in (\ref{j2}) becomes%
\begin{equation}
\partial _{k}D^{k}\gamma _{0}=[c^{-1}j^{(C)0}-c^{-2}\varepsilon
^{kij0}\partial _{k}M_{i}U_{j}]\gamma _{0},  \label{kd}
\end{equation}%
whereas the part with $\gamma _{i}$ is
\begin{equation}
\varepsilon ^{ijk0}\partial _{j}H_{k}\gamma _{i}=[j^{(C)i}+c\partial
_{0}D^{i}+((U^{k}\partial _{k})P^{i}-U^{i}(\partial
_{k}P^{k}))+c^{-1}\varepsilon ^{ijk0}\partial _{0}M_{j}U_{k}]\gamma _{i}.
\label{kh}
\end{equation}

The equation (\ref{j2}), i.e., Eqs. (\ref{kd}) and (\ref{kh}), can be
compared with the usual form of Maxwell's equations for moving media, which
deal with 3D vectors, e.g., [15, 28, 29], etc. Again, as in Sec. 4, the
comparison will be possible \emph{only} if the term with $\gamma _{0}$ is
considered separately from those ones with $\gamma _{i}$, as in (\ref{e})
and (\ref{gi}) and if in Eq. (\ref{j2}) \emph{only the components are taken
into account. }As before, we argue that \emph{from the viewpoint of the
geometric approach, }i.e., \emph{the ISR, such a procedure is not correct in}
\emph{4D spacetime}. If, as in Sec. 4, in Eq. (\ref{j2}), i.e., in $%
a^{0}\gamma _{0}+a^{i}\gamma _{i}=0$, one takes that $a^{0}=0$, and multiply
the spatial components of $E$, $P$, $M$ and $j^{(C)}$ from $a^{0}$ by \emph{%
the unit 3D vectors} $\mathbf{i}$, $\mathbf{j}$, $\mathbf{k}$ then the term
with $\gamma _{0}$ will become the equation

\begin{equation}
\nabla \cdot \varepsilon _{0}\mathbf{E}(\mathbf{r},t)=\rho ^{(C)}(\mathbf{r}%
,t)-\nabla \cdot \lbrack \mathbf{P}(\mathbf{r},t)-c^{-2}(\mathbf{M}(\mathbf{r%
},t)\times \mathbf{U})].  \label{elv}
\end{equation}%
Usually, as, e.g., in [28] (the derivation of Eqs. (9-18) (1-4)) the case
of: \textquotedblleft a non-magnetized medium moving with a velocity $%
\mathbf{u}$ which is small compared with velocity of
light,\textquotedblright\ is considered. This means that in (\ref{j1}) one
has to take not only $\beta _{u}\ll 1$, which leads to (\ref{j2}), but also $%
M_{i}=0$. Then, instead of the part with $\gamma _{0}$ from (\ref{j2}),
i.e., Eq. (\ref{kd}), one gets the equation $\partial _{k}D^{k}\gamma
_{0}=\rho ^{(C)}\gamma _{0}$. In the formulation with 3D vectors that
equation corresponds to, e.g., Eq. (9-18)(1) in [28], or Eq. (7.55)(i) in
[26], $\nabla \cdot \mathbf{D}=\rho ^{(C)}$.

In the same way, the terms with $\gamma _{i}$ from (\ref{j2}), i.e., Eq. (%
\ref{kh}), can be compared with the usual form of Maxwell's equations for
moving media, which deals with 3D vectors, e.g., Eq. (9-18) (4) in [28], or
the equations in Problem 6.8 in [29]. Then, the terms with $\gamma _{i}$
from (\ref{j2}) correspond to the following equation with 3D vectors
\begin{eqnarray}
\nabla \times \mathbf{B}(\mathbf{r},t) &=&\mu _{0}[\mathbf{j}^{(C)}(\mathbf{r%
},t)+\partial (\varepsilon _{0}\mathbf{E}(\mathbf{r},t)+\mathbf{P(\mathbf{r},%
}t\mathbf{))}/\partial t+\nabla \times (\mathbf{P}(\mathbf{r},t)\times
\mathbf{U})  \notag \\
&&+(1/c^{2})\partial (\mathbf{U}\times \mathbf{M(\mathbf{r},}t\mathbf{))/}%
\partial t\mathbf{+}\nabla \times \mathbf{M}(\mathbf{r},t)],  \label{blv}
\end{eqnarray}%
or, from (\ref{kh}), the equation (\ref{blv}) can be written in the
following form
\begin{equation}
\nabla \times \mathbf{H}(\mathbf{r},t)=\mathbf{j}^{(C)}(\mathbf{r}%
,t)+\partial \mathbf{D}(\mathbf{r},t)\mathbf{/}\partial t+(1/c^{2})\partial (%
\mathbf{U}\times \mathbf{M(\mathbf{r},}t\mathbf{))/}\partial t+\nabla \times
(\mathbf{P}(\mathbf{r},t)\times \mathbf{U}),  \label{hd}
\end{equation}%
where the 3D vectors $\mathbf{D}=\varepsilon _{0}\mathbf{E}+\mathbf{P}$ and $%
\mathbf{H}=(1/\mu _{0})\mathbf{B}-\mathbf{M}$ have been introduced. Taking
in Eq. (\ref{j2}) that not only $\beta _{u}\ll 1$ but that $M_{i}=0$ as
well, i.e., that a non-magnetized medium is considered, then instead of Eq. (%
\ref{hd}) we find
\begin{equation}
\nabla \times \mathbf{B}(\mathbf{r},t)=\mu _{0}[\mathbf{j}^{(C)}(\mathbf{r}%
,t)+\partial \mathbf{D}(\mathbf{r},t)\mathbf{/}\partial t+\nabla \times (%
\mathbf{P}(\mathbf{r},t)\times \mathbf{U})].  \label{bp}
\end{equation}%
This equation is the fourth equation in Problem 6.8 in [29]. It differs from
Eq. (9-18) (4) in [28], which contains an additional term $\mu _{0}\rho
^{(C)}\mathbf{U}$. As seen from (\ref{bp}) the appearance of that additional
term is not justified.

The whole consideration on the difference between Maxwell's equations with
3D vectors and the equations with 4D quantities that is presented at the end
of Sec. 4 holds in the same measure here. However, the difference between
these two approaches (3D vectors versus 4D geometric quantities) is even
bigger for the case examined in this section. Namely, due to the existence
of the terms with $P^{0}$ and $M^{0}$ in (\ref{j1}) that equation \emph{%
cannot} be compared with Maxwell's equations with 3D vectors. The comparison
can be made only for low velocities of the medium when Eq. (\ref{j1})
reduces to Eq. (\ref{j2}).

There is also an additional difference between Maxwell's equations with 3D
vectors, e.g., Eqs. (9-18) in [28], and our Eqs. (\ref{j1}) and (\ref{j2}).
It is stated in [28] (under Eqs. (9-18)): \textquotedblleft Note that
Maxwell's equations for moving (nonmagnetic) media in the form given by Eq.
(9-18) (4) are \textquotedblleft mixed,\textquotedblright\ i.e., the sources
$\mathbf{j}_{true}$, $\mathbf{P}$, $\rho _{true}$, are measured in the
moving medium, while the fields are given in the stationary
frame.\textquotedblright\ On the other hand, as already stated above, \emph{%
all quantities} in (\ref{j1}) and (\ref{j2}) are determined in the
laboratory frame, which is the $\gamma _{0}$-frame. Moreover, as already
explained, all quantities in (\ref{j1}) and (\ref{j2}) are correctly defined
in 4D spacetime and they correctly transform under the LT, which is not the
case with Maxwell's equations with 3D vectors.\bigskip \bigskip

\noindent \textbf{6. Comparison with Galilean Electromagnetism\bigskip }

\noindent At this place it is worth mentioning that in a recent review [30]
under the title \textquotedblleft Forty years of Galilean Electromagnetism
(1973 - 2013)\textquotedblright\ and in the references therein, e.g., [31,
32], it is argued that, [30], \textquotedblleft Galilean Electromagnetism is
precisely the low-velocity limit of Special Relativity when applied to
Classical Electromagnetism.\textquotedblright\ There, it is also stated that
in a Galilean limit the usual Maxwell-Minkowski equations with the 3D
vectors $\mathbf{E}$, $\mathbf{B}$, $\mathbf{D}$ and $\mathbf{H}$, Eq. (1)
in [30], have to be replaced by two Galilean limits, the magnetic and
electric limits, i.e., with two sets of low-velocity formulae, Eqs. (16) and
(18) in [30], respectively. Furthermore, in [30], it is considered that the
UT, Eqs. (\ref{JCB}) and (\ref{ps}) here and the similar ones for $\mathbf{D}
$ and $\mathbf{H}$, Eq. (3) in [30] or in [31], are the relativistically
correct LT, but that in Galilean approximation they have to be replaced by
two sets of low-velocity formulae, the magnetic and electric limits, which
are presented, e.g., in Sec. 3 in [30]. The same happens with Minkowski's
constitutive equations, Eq. (9), i.e., Eq. (10) in [30] or in [31], which
are replaced by the Galilean magnetic constitutive equations, Eq. (13) in
[30], and the Galilean electric constitutive equations, Eq. (14) in [30]. In
addition, it is considered in [30], as in almost all other usual approaches,
that the contraction of lengths and the dilation of time are
\textquotedblleft the phenomena inherent to Special
Relativity.\textquotedblright

However, in 4D spacetime the physical quantities are represented in a
mathematically correct way by 4D geometric quantities that properly
transform under the LT, e.g., (\ref{ELT}), and not by 3D quantities that
transform by the UT, (\ref{JCB}) and (\ref{ps}) and the similar UT for $%
\mathbf{D}$ and $\mathbf{H}$, Eq. (3) in [30] or in [31]. As discussed in
Sec. 4, according the UT, (\ref{JCB}) and (\ref{ps}), i.e., Eq. (3) in [30]
or in [31], e.g., the transformed $\mathbf{E}^{\prime }$ is expressed by the
mixture of the 3D vectors $\mathbf{E}$ and $\mathbf{B}$. Using that
essential feature of the UT of the fields as 3D vectors a Galilean limit of
the field transformations is derived in two steps in, e.g., [32]. First, the
quasi-static approximation, $\beta \ll 1$, is taken and then the assumption
on the relative magnitude of $\left\vert \mathbf{E}\right\vert $ and c$%
\left\vert \mathbf{B}\right\vert $ is taken into account. If the magnetic
field is dominant then the magnetic limit, Eq. (7) in [32] is obtained,
whereas in the opposite case the electric limit, Eq. (8) in [32], is
obtained. On the other hand, as discussed in Sec. 2, the essential feature
of the mathematically correct LT of 4D fields, like (\ref{ELT}), is that the
LT transform, e.g., the electric field vector only to the electric field
vector; there is no mixing with the magnetic field vector. This means that
both Galilean limits, the magnetic limit and the electric limit of the field
transformations, Eqs. (7) and (8) in [32] are meaningless in 4D spacetime in
which the fields are represented by 4D geometric quantities that\ correctly
transform under the LT.

In [5] the constitutive relations are formulated in terms of coordinate-free
quantities that correctly transform under the LT. First, in Sec. 3 they are
formulated as the relations between $\mathcal{M}$ and $F$, Eqs. (11) and
(12) in [5]. Then, using the decompositions of $F$ (\ref{E2}) and $\mathcal{M%
}$ (\ref{M1}) the basic constitutive relations for $P(x)$ and $M(x)$, Eqs.
(13) and (14) in [5], are obtained. They show how $P(x)$ and $M(x)$ depend
on $E$, $B$ and two different velocity vectors, $u$ and $v$. These
constitutive relations differ from all previous expressions and they are not
reported in any previous approach. In Secs. 5 - 5.2 in [5] it is explained
that Minkowski's constitutive relations, Eqs. (23) and (24), or (25) in [5],
i.e., Eq. (11) in [30], or Eq. (10) in [31], are not the \emph{relativistic}
constitutive equations. They are the relations with 3D vectors that
transform according to the UT, like (\ref{JCB}) and (\ref{ps}), and these
transformations, as already stated several times, are not the LT. Therefore,
the constitutive relations, Eqs. (13) and Eq. (14) in [30] are not any kind
of a quasi-static approximation of the relativistically correct constitutive
relations. As stated in [5], there is only one mathematically correct
quasi-static approximation of the constitutive relations for $P(x)$ and $%
M(x) $, which is given by Eqs. (19) and (21) in [5] in which for the low
velocities of the medium it is only taken that $\beta _{u}\ll 1$, i.e., $%
\gamma _{u}\simeq 1$, see Secs. 4, 5 - 5.2 in [5].

It can be concluded from the preceding discussion that both Galilean limits,
the Galilean magnetic Maxwell - Minkowski equations and the Galilean
electric Maxwell - Minkowski equations, e.g., Eqs. (16) and (18) in [30],
respectively, are ill-defined in 4D spacetime; they are not a quasi-static
approximation of the relativistically correct field equations. As shown in
Sec. 4 and in this section the usual Maxwell-Minkowski equations with the 3D
vectors $\mathbf{E}$, $\mathbf{B}$, $\mathbf{D}$ and $\mathbf{H}$, Eq. (1)
in [30], \emph{are not} the relativistic form of Maxwell's equations,
because in 4D spacetime there is no room for 3D vectors and their UT. The
mathematically correct field equation is, e.g., Eq. (\ref{j1}) and the
equations from which (\ref{j1}) is derived and there is only one its
quasi-static approximation, i.e., the approximation for the case of low
velocities of the medium, it is Eq. (\ref{j2}).

Furthermore, as discussed in Sec. 4, it is exactly proved in [27] and in
[23] that the relativity of simultaneity, the Lorentz contraction and the
time dilation \emph{are ill-defined in 4D spacetime}, because they depend on
the chosen synchronization. Hence, they are not, as argued in [30]
\textquotedblleft the phenomena inherent to Special
Relativity.\textquotedblright\ This dependence on chosen synchronization
holds in the same measure for the usual Maxwell-Minkowski equations, Eq. (1)
in [30], for the UT of the fields as 3D vectors, Eq. (3) in [30] or in [31],
for Minkowski's constitutive relations, Eq. (11) in [30] and for the whole
Galilean Electromagnetism including the comparison with experiments that is
presented in Secs. 5 - 8.2 in [30].\bigskip \bigskip

\noindent \textbf{7. Motional electromotive force in the approaches with 3D }

\textbf{quantities and with 4D geometric quantities\bigskip }

\noindent On the other hand, as already mentioned in Sec. 4, the approach
with 4D geometric quantities and their LT, always agrees with the principle
of relativity and it is in a true agreement with all experiments in
electromagnetism and all experiments that test SR, see [17, 18, 2, 25, 27].
In Secs. 7.1 and 7.2, instead of to discuss the experiments which are
presented in Secs. 5 - 8.2 in [30], we shall briefly present the discussion
of the motional electromotive force (emf) that is exposed in Secs. 5 - 5.2
in [17]. However, in Sec. 7.2, some changes relative to Sec. 5.2 in [17]
will be introduced ($F$ instead of $E$ and $B$). It is a nice example that
illustrates the fundamental difference between the LT, e.g., (\ref{ELT}),
and the UT (\ref{JCB}) and (\ref{ps}), i.e., between the approach with 4D
geometric quantities and the usual approach with 3D vectors.\bigskip

\noindent \textit{7.1. Motional emf with 3D quantities}\textbf{\bigskip }

\noindent The motional emf is produced in an electrical circuit when a
circuit or part of a circuit moves in a magnetic field. In Sec. 5.1 in [17]
the emf $\varepsilon $ is calculated using 3D quantities, the 3D Lorentz
force $\mathbf{F}_{L}=q\mathbf{E}+q\mathbf{U}\times \mathbf{B}$, the 3D $%
\mathbf{E}$ and $\mathbf{B}$ and their UT (\ref{JCB}) as in Secs. 6.4 in
[29], 7.2 in [33], 9-5 in [28], 5.6 in [15] and in all other calculations in
the usual approaches. The emf $\varepsilon $ of a complete circuit is
defined by means of the Lorentz force $\mathbf{F}_{L}$ that acts on a charge
$q$, which is at rest relative to the section $\mathbf{dl}$ of the circuit,
\begin{equation}
\varepsilon =\oint (\mathbf{F}_{L}\mathbf{/}q)\cdot \mathbf{dl,}  \label{eps}
\end{equation}%
Eq. (26) in [17]. The important remark is that it is implicitly assumed in
these equations for $\varepsilon $ that the integral is taken over the whole
circuit at the same moment of time in $S$, say $t=0$. Let us take that in
the laboratory frame $S$ a conducting bar is moving in a steady uniform
magnetic field (3-vector) $\mathbf{B=-}B\mathbf{k}$ with velocity 3-vector $%
\mathbf{U}$ parallel to the $x$ axis. The length of the bar is $l$ and it
moves parallel to the $y$ axis. There is no external applied electric field
in $S$, $\mathbf{E=}0$. Since in $S$ $\mathbf{E=}0$ and the components of $%
\mathbf{B}$ are $\mathbf{(}0,0\mathbf{,-}B\mathbf{)}$\ the emf $\varepsilon $
is determined by the contribution of the magnetic part of $\mathbf{F}_{L}$,
i.e., $q\mathbf{U}\times \mathbf{B}$, as $\varepsilon =\int_{o}^{l}UBdy=UBl$%
, which is Eq. (27) in [17].

In $S^{\prime }$ the conducting bar is at rest. Then, \emph{according to the
UT }(\ref{JCB})\emph{\ }of the 3D $\mathbf{E}$ and $\mathbf{B}$ the observer
in the $S^{\prime }$ frame \textquotedblleft sees\textquotedblright\ $%
E_{y}^{\prime }=\gamma UB$ and $B_{z}^{\prime }=-\gamma B$. Hence, \emph{in}
$S^{\prime }$, there is not only the magnetic field but \emph{an electric
field} $\mathbf{E}^{\prime }$ as well. The contribution of the magnetic part
(due to $B_{z}^{\prime }$) of the Lorentz force $\mathbf{F}_{L}^{\prime }$
to the emf $\varepsilon ^{\prime }$ is zero and only the contribution of the
electric part (due to $E_{y}^{\prime }$) of the Lorentz force remains, which
is $\varepsilon ^{\prime }=\int_{o}^{l}\gamma UBdy=\gamma UBl$, Eq. (29) in
[17]. Observe that the integral is taken at the same moment of time $%
t^{\prime }$ in $S^{\prime }$, which can be arbitrarily chosen, say $%
t^{\prime }=0$, or $t^{\prime }=10s$, ... . The moments of time $t$ in $S$
and $t^{\prime }$ in $S^{\prime }$ are not connected in any way. The LT
cannot transform the moment of time $t$ in $S$ again, exclusively, to some $%
t^{\prime }$ in $S^{\prime }$. According to the LT, to one $t$ in $S$ will
correspond many $t^{\prime }$ in $S^{\prime }$ depending on the spatial
position in $S^{\prime }$; $t=\gamma (t^{\prime }+Ux^{\prime }/c^{2})$. This
remark clearly shows that the usual definition of $\varepsilon $, Eq. (\ref%
{eps}), is not relativistically correct definition. Obviously, \emph{the emf
}$\varepsilon ^{\prime }$ \emph{in }$S^{\prime }$ \emph{is not equal to the
emf} $\varepsilon $, \emph{determined} \emph{in} $S$;
\begin{equation}
\varepsilon =UBl,\ \varepsilon ^{\prime }=\gamma UBl,\quad \varepsilon
^{\prime }\neq \varepsilon .  \label{eec}
\end{equation}%
Consequently, \emph{the principle of relativity is not satisfied; }the emf
obtained by the application of the UT (\ref{JCB})\ is different for
relatively moving 4D observers. This explicitly shows that the conventional
calculation of $\varepsilon $ and the UT (\ref{JCB}) of the 3D $\mathbf{E}$
and $\mathbf{B}$ are not relativistically correct, i.e., the UT (\ref{JCB})
are not the LT. The fact that $\varepsilon $ and $\varepsilon ^{\prime }$ do
not significantly differ for low velocities, $U\ll c$, is completely
irrelevant; the principle of relativity is not satisfied in the usual
approach. Of course, the same holds for the Galilean
Electromagnetism.\bigskip

\noindent \textit{7.2. Motional emf with 4D geometric quantities}\textbf{%
\bigskip }

\noindent In Sec. 5.2 in [17] the emf $\varepsilon $ is calculated using 4D
geometric quantities, the vectors $E$ and $B$, the vector of the Lorentz
force $K$, etc. There, it is found that in the approach with 4D geometric
quantities and their LT, e.g., (\ref{ELT}), the same value for $\varepsilon $
is always obtained, $\varepsilon =\gamma UBl$, which means that the
principle of relativity is naturally satisfied. Here, we shall obtain the
same result as in Sec. 5.2 in [17] but dealing with $F$ and not with its
decompositions (\ref{E2}) and (\ref{fm}). The Lorentz force $K$ is defined
as an abstract vector and as a CBGQ in the standard basis by the following
relations

\begin{equation}
K=(q/c)F\cdot u,\quad K=K^{\mu }\gamma _{\mu }=(q/c)(F^{\mu \nu }u_{\nu
})\gamma _{\mu }.  \label{KEB}
\end{equation}%
where $u$ is the velocity vector of the considered charge. \emph{The emf} $%
\varepsilon $ \emph{is defined as an invariant 4D quantity, the Lorentz
scalar},

\begin{equation}
\varepsilon =\int_{\Gamma }(K/q)\cdot dl,\quad \varepsilon =\int_{\Gamma
}(K^{\mu }/q)dl_{\mu }=(1/c)\int_{\Gamma }F^{\mu \nu }u_{\nu }dl_{\mu },
\label{emfi}
\end{equation}%
where \emph{vector} $dl$ is the infinitesimal spacetime length and $\Gamma $
is the spacetime curve.

In the laboratory frame $S$ with the standard basis in it the components of $%
u$ and $dl$ are $u^{\mu }=(\gamma c,\gamma U,0,0)$, $dl^{\mu
}=(0,0,dl^{2}=dy,0)$. It can be seen from Eq. (\ref{ebv}) that in the
considered case all components $F^{\mu \nu }$\ are zero except $F^{21}$\ ($%
F^{12}$), which is $F^{21}=cB^{3}$. This result and Eq. (\ref{KEB}) yield
that the components of the Lorentz force $K$ are $K^{0}=K^{1}=K^{3}=0$, but $%
K^{2}=(q/c)F^{21}(-\gamma U)$. Hence, the emf $\varepsilon $ is%
\begin{equation}
\varepsilon =(1/c)\int_{0}^{l}F^{21}(-\gamma U)dy=(1/c)F^{21}(-\gamma U)l.
\label{el}
\end{equation}%
This result can be compared with that one from Sec. 5.2 in [17], $%
K^{2}=\gamma qUB$ and $\varepsilon =\gamma UBl$, using the relations $%
F^{21}=cB^{3}$, and $B^{3}=B_{z}=-B$. Then%
\begin{equation}
\varepsilon =(1/c)\int_{0}^{l}F^{21}(-\gamma U)dy=\gamma UBl,  \label{el1}
\end{equation}%
which is the same as in [17]. In contrast to the usual approaches with 3D
quantities the expression for $\varepsilon $ (\ref{emfi}) is independent of
the chosen reference frame and of the chosen basis in it; $\varepsilon $ is
the same in $S$ and in the relatively moving $S^{\prime }$ frame,
\begin{equation}
\varepsilon =\int_{\Gamma }(K^{\mu }/q)dl_{\mu }=\int_{\Gamma }(K^{\prime
\mu }/q)dl_{\mu }^{\prime }=\gamma UBl,  \label{el2}
\end{equation}%
which is Eq. (37) in [17]. This result for $\varepsilon $ can be checked
directly performing the LT of all vectors from $S$ to $S^{\prime }$. Observe
that in $S^{\prime }$ the 3-velocity $\mathbf{U}$ is zero, but the velocity
vector $u$ \emph{is not}, $u=c\gamma _{0}$, i.e., $u^{\prime \mu }=(c,0,0,0)$%
. The same value for $\varepsilon $ will be obtained if another basis, e.g.,
the $\{r_{\mu }\}$ basis, will be used in both frames. Obviously, \emph{in
the 4D geometric approach, the principle of relativity is naturally
satisfied.}

The result that the conventional theory with the 3D $\mathbf{E}$ and $%
\mathbf{B}$ and their UT (\ref{JCB}) yields different values for $%
\varepsilon $ for relatively moving inertial observers, $\varepsilon =UBl$
in $S$ and $\varepsilon ^{\prime }=\gamma UBl$ in $S^{\prime }$, whereas
ISR, i.e., the approach with 4D geometric quantities and their LT, e.g., (%
\ref{ELT}), yields always the same value for $\varepsilon $, $\varepsilon
=\gamma UBl$, Eq. (\ref{el2}), is very strong evidence that the usual
approach is not relativistically correct. It is for the experimentalists to
find the way to measure the emf $\varepsilon $ with a great precision and to
see that in the laboratory frame $\varepsilon =\gamma UBl$ and not simply $%
\varepsilon =UBl$.

Observe that in this calculation with 4D geometric quantities all quantities
are invariant under the passive LT, e.g., $u=u^{\nu }\gamma _{\nu
}=u^{\prime \nu }\gamma _{\nu }^{\prime }=u_{r}^{\nu }r_{\nu }=u_{r}^{\prime
\nu }r_{\nu }^{\prime }$, $K=K^{\nu }\gamma _{\nu }=K^{\prime \nu }\gamma
_{\nu }^{\prime }=K_{r}^{\nu }r_{\nu }=K_{r}^{\prime \nu }r_{\nu }^{\prime }$%
, etc., which means that the observers in $S$ and $S^{\prime }$ are
\textquotedblleft looking\textquotedblright\ at the same quantity, for
example, the Lorentz force $K$. This consideration shows that ISR, i.e., the
approach with 4D geometric quantities and their LT (\ref{ELT}) is
substantially different than the usual approaches with the 3D quantities and
their UT (\ref{JCB}). The former is completely suited to the symmetry of 4D
spacetime, which is not the case with the latter.

There are many similar examples in the literature; it will be always found
that there is a fundamental difference between the UT of the 3D $\mathbf{E}$
and $\mathbf{B}$ (\ref{JCB}) and the LT of 4D geometric quantities
representing the electric and magnetic fields, e.g., (\ref{ELT}), and that
the 4D geometric approach, i.e., ISR, correctly describes electromagnetic
phenomena in all relatively moving 4D inertial frames of reference. One
important experiment, the Faraday disk, which leads to the same conclusions,
is considered in detail in [18].\bigskip \bigskip

\noindent \textbf{8. Discussion and conclusions}\bigskip

\noindent There are several important differences between the field
equations reported here and all others in previous literature including the
modern textbook on classical electrodynamics [34], which uses the calculus
of exterior forms.

First, instead of dealing with the electromagnetic excitation $\mathcal{H}$ (%
\ref{F5}) and the field equation with it (\ref{F6}) we exclusively deal with
the equations (\ref{F4}) for the electromagnetic field $F$ and a matter
field $\mathcal{M}$ as the primary equations for electromagnetism in moving
media. As discussed in Secs. 2 and 3, the expression for $\mathcal{H}$ (\ref%
{F5}) in terms of $F$ and $\mathcal{M}$ is in some sense unsatisfactory,
since $F$ and $\mathcal{M}$ are physically different kind of entities.
Furthermore, what is particularly important, in general, two different
velocity vectors, $v$ - the velocity of the observers and $u$ - the velocity
of the moving medium, enter into the decompositions of $F$ and $\mathcal{M}$%
, Eqs. (\ref{E2}) and (\ref{M1}), respectively. For this reason we also do
not deal with the decomposition of $\mathcal{H}$ (\ref{h1}) into the
electric and magnetic excitations $D$ and $H$, respectively, where $%
D=\varepsilon _{0}E+P$ and $H=(1/\mu _{0})B-M$. As stated in Sec. 3, such a
decomposition as (\ref{h1}) is possible if only one velocity, the velocity
of the medium $u$, is taken into account, or the case $u=v$ is considered,
or both decompositions (\ref{E2}) and (\ref{M1}) are made with the same
velocity vector, either $u$ or $v$. Recently, the last case is considered in
[35], but with $F$ and $\mathcal{H}$.

The second important difference refers to the interpretation of the field
equations. The basic field equation (\ref{F7}) contains two different
velocities $u$ and $v$. It is also written as Eqs. (\ref{A1}) and (\ref{A2}%
), i.e., with CBGQs, (\ref{I1}) and (\ref{I2}), respectively. From these
equations it is visible that \emph{in 4D spacetime}, in contrast to the
formulation of electromagnetism in terms of Maxwell's equations with the 3D
vectors $\mathbf{E}$, $\mathbf{B}$, $\mathbf{P}$ and $\mathbf{M}$, \emph{%
there are no two laws, the Amp\`{e}re-Maxwell law and Gauss's law, but only
one law, that is expressed by Eq. }(\ref{A1}) ((\ref{I1})),\emph{\ }i.e.,%
\emph{\ Eq.} (\ref{Ej}) ((\ref{J})),\emph{\ and the same for Eq. }(\ref{A2})
((\ref{I2})) and \emph{Faraday's law and the law that expresses the absence
of magnetic charge. }

Furthermore, the interesting results are obtained in Secs. 4 and 5. There,
the field equations, written in the standard basis (\ref{I1}), i.e., (\ref{J}%
), and (\ref{I2}), are compared with the usual form (with 3D vectors) of
Maxwell's equations for moving media. In Sec. 4, it is shown that the
comparison is possible \emph{only} if the term with $\gamma _{0}$ is
considered \emph{separately} from those ones with $\gamma _{i}$ and if in
Eqs. (\ref{e}) and (\ref{gi}) \emph{only the components} are taken into
account. In order to get the usual equations with 3D vectors, (\ref{p4}),(%
\ref{p5}), (\ref{g2}), (\ref{g3}) and (\ref{dh}), these components have to
be multiplied by \emph{the unit 3D vectors} $\mathbf{i}$, $\mathbf{j}$, $%
\mathbf{k}$. Moreover, as shown in Sec. 5, such a procedure is not
applicable to Eq. (\ref{j1}), but only to Eq. (\ref{j2}), which is derived
from (\ref{j1}) for the case of low velocities of the medium. As explained
in Sec. 4, \emph{the above mentioned steps in the comparison are not
mathematically correct in 4D spacetime}. Hence, in 4D spacetime, the
equations (\ref{F7}), (\ref{A1}), (\ref{A2}), (\ref{I1}), (\ref{I2}), ... ,
with the 4D geometric quantities $E$, $B$, $P$ and $M$ that correctly
transform under the LT, e.g., for the components (\ref{ELT}), or for the
vector $E$, Eq. (8) in [5], i.e., Eq. (13) in [11], \emph{are not equivalent
}to the usual Maxwell equations (\ref{p4}), (\ref{p5}), (\ref{g2}), (\ref{g3}%
), (\ref{dh}), (\ref{elv}), (\ref{blv}), ... , with the 3D vectors $\mathbf{E%
}$, $\mathbf{B}$, $\mathbf{P}$ and $\mathbf{M}$ that transform according to
UT, e.g., Eqs. (\ref{JCB}) and (\ref{ps}).

It will be important for physics to examine the theoretical and experimental
consequences of the results that are obtained in this paper. An interesting
consequence that can be experimentally examined is already mentioned in
connection with Eq. (\ref{Ej}). There, it is stated that if we have a
magnetization $M$, a permanent magnet, moving \emph{or stationary}, but with
a negligible permanent polarization $P$ and without $j^{(C)}$, then, as can
be seen from (\ref{Ej}), $M$ will induce \emph{both} $B$ \emph{and} $E$. In
Sec. 8 in [22] the existence of the electric field from a stationary
permanent magnet is investigated in detail and that field is used in the
consideration of the \textquotedblleft charge-magnet
paradox\textquotedblright\ in [22] and [36].

At this place it is worth mentioning a recent paper, [37], under the title
\textquotedblleft Nature of Electric and Magnetic Fields; How the Fields
Transform.\textquotedblright\ In that paper, in 3.1 and 3.3 the
mathematically correct proofs are given that the electric and magnetic
fields, $E(x)$ and $B(x)$, respectively are properly defined vectors on 4D
spacetime and not the usual 3D vectors $\mathbf{E}$ and $\mathbf{B}$.
Furthermore, it is proved in [37] that the correct LT of the electric field
are given by (\ref{ELT}) and not by the UT of the 3D vectors Eqs. (11.148)
and (11.149) in [1]. The proof 3.3 from [37] is based on the mathematical
theorem presented in Sec. 3 here, i.e., it is given by the relations (\ref%
{E2}) - (\ref{ebv}) here. The proof 3.1 is a simple but very strong
mathematical argument, which is stated by Oziewicz, e.g., in [38]: \emph{%
What is essential for the number of components of a vector field is the
number of variables on which that vector field depends, i.e.,} \emph{the
dimension of its domain}. In general, the dimension of a vector field that
is defined on a n-dimensional space is equal - n. \emph{The electric and
magnetic fields are defined on a 4D space, i.e., the spacetime. They are
always functions of the position vector} $x$. \emph{This means that they are
not the usual 3D fields,\ but they are properly defined vectors on 4D
spacetime}, $E(x)$ and $B(x)$. The same holds for the polarization vector $%
P(x)$ and the magnetization vector $M(x)$. Since $E$, $B$, $P$ and $M$ are
4D vectors they must transform \emph{under the LT as any other vector
transforms},\emph{\ }e.g., \emph{the electric field vector must transform
again} \emph{to the electric field vector} like in (\ref{ELT}). In [39] the
same result is obtained for the electric field as a bivector and for the
magnetic field as well.\bigskip \bigskip

\noindent \textbf{Acknowledgments\bigskip }

\noindent It is a pleasure to acknowledge to Larry Horwitz and Martin Land
for inviting me to the IARD conferences and for their continuos support of
my work. I am also very grateful to Zbigniew Oziewicz for numerous and very
useful discussions during years and to him and to Alex Gersten for the
continuos support of my work.\bigskip \bigskip

\noindent \textbf{References\bigskip }

\noindent \lbrack 1] Minkowski H 1908 \textit{Nachr. Ges. Wiss. G\"{o}ttingen%
} 53;

Reprinted in: 1910 \textit{Math. Ann.} \textbf{68} 472;

English translation in: M N Saha and S N Bose 1920 \textit{The Principle }

\textit{of Relativity: Original Papers by A. Einstein and H. Minkowski}

(Calcutta: Calcutta University Press)

\noindent \lbrack 2] Ivezi\'{c} T 2005 \textit{Found. Phys.} \textit{Lett.}
\textbf{18} 401

\noindent \lbrack 3] Einstein A 1905 \textit{Annalen der Physik} \textbf{17}
891; English translation in:

Perrett W and Jeffery G B 1952 \textit{The Principle of Relativity}

(New York: Dover)

\noindent \lbrack 4] Hestenes D 1966 \textit{Space-Time Algebra (}New York:
Gordon \& Breach);

Hestenes D and Sobczyk G 1984 \textit{Clifford Algebra to }

\textit{Geometric Calculus }(Dordrecht: Reidel);

Doran C and Lasenby A 2003 \textit{Geometric algebra for physicists}

(Cambridge: Cambridge University Press)

\noindent \lbrack 5] Ivezi\'{c} T 2012 \textit{Int. J. Mod. Phys. B} \textbf{%
26} 1250040

\noindent \lbrack 6] Jackson J D 1998 \textit{Classical Electrodynamics} 3rd
edn (New York: Wiley)

\noindent \lbrack 7] Hehl F W 2008 \textit{Annalen der Physik} \textbf{17}
691

\noindent \lbrack 8] Obukhov Y N 2008 \textit{Annalen der Physik} \textbf{17}
830

\noindent \lbrack 9] Itin Y and Friedman Y 2008 \textit{Annalen der Physik}
\textbf{17} 769

\noindent \lbrack 10] Ludvigsen M 1999 \textit{General\ Relativity}, \textit{%
A Geometric Approach}

(Cambridge: Cambridge University Press);

Sonego S and Abramowicz M A J 1998 \textit{J. Math. Phys.} \textbf{39} 3158

\noindent \lbrack 11] Ivezi\'{c} T 2010 \textit{Phys. Scr.} \textbf{82}
055007

\noindent \lbrack 12] M\o ller C 1972 \textit{The Theory of Relativity} 2nd
edn (Oxford: Clarendon Press)

\noindent \lbrack 13] Hillion P 1993 \textit{Phys. Rev. E} \textbf{48} 3060;

\noindent \lbrack 14] Dereli T, Gratus J and Tucker R W 2007 \textit{Phys.
Lett. A} \textbf{361} 190

\noindent \lbrack 15] Bladel J Van 1984 \textit{Relativity and Engineering}
(Berlin: Springer-Verlag)

\noindent \lbrack 16] Ivezi\'{c} T 2003 \textit{Found. Phys.} \textbf{33}
1339

\noindent \lbrack 17] Ivezi\'{c} T 2005 \textit{Found. Phys.} Lett. \textbf{%
18 }301

\noindent \lbrack 18] Ivezi\'{c} T 2005 \textit{Found. Phys.} \textbf{35}
1585

\noindent \lbrack 19] Ivezi\'{c} T 2008 \textit{Fizika A }\textbf{17} 1

\noindent \lbrack 20] Ivezi\'{c} T 2008 \textit{arXiv: }0809.5277

\noindent \lbrack 21] Ivezi\'{c} T 2007 \textit{Phys. Rev. Lett.} \textbf{98
}108901

\noindent \lbrack 22] Ivezi\'{c} T 2013 \textit{J. Phys.: Conf. Ser.}
\textbf{437} 012014

\noindent \lbrack 23] Ivezi\'{c} T 2001 \textit{Found. Phys.} \textbf{31}
1139

\noindent \lbrack 24] Ivezi\'{c} T 2010 \textit{Phys. Scr. }\textbf{81}
025001

\noindent \lbrack 25] Ivezi\'{c} T 2007 \textit{Found. Phys.} \textbf{37} 747

\noindent \lbrack 26] Griffiths D J 2013 \textit{Introduction to
Electrodynamics} 4th edn

(Boston: Pearson)

\noindent \lbrack 27] Ivezi\'{c} T 2002 \textit{Found. Phys. Lett.} \textbf{%
15} 27;

Ivezi\'{c} T 2001 \textit{Preprint} physics/0103026;

Ivezi\'{c} T 2001 \textit{Preprint} physics/0101091

\noindent \lbrack 28] Panofsky W K H and Phillips M 1962 \textit{Classical
electricity and magnetism}

2nd edn (Reading: Addison-Wesley)

\noindent \lbrack 29] Rosser W G W 1968 \textit{Classical Electromagnetism
via Relativity}

(New York: Plenum)

\noindent \lbrack 30] Rousseaux G 2013 \textit{Eur. Phys. J. Plus} \textbf{%
128} 81

\noindent \lbrack 31] Rousseaux G 2008 \textit{Europhys. Lett.} \textbf{84}
20002

\noindent \lbrack 32] de Montigny M and Rousseaux G 2006 \textit{Eur. Phys.
J. Plus} \textbf{27} 755

\noindent \lbrack 33] Purcell E M 1985 \textit{Electricity and Magnetism }%
2nd edn (New York:

McGraw-Hill)

\noindent \lbrack 34] Hehl F W and Obukhov Yu N 2003 \textit{Foundations of
Classical }

\textit{Electrodynamics: Charge, flux, and metric} (Boston: Birkh\"{a}user)

\noindent \lbrack 35] Goto Shin-itiro, Tucker R W and Walton T J 2011
\textit{Proc. Roy. Soc. A}

\textbf{467} 59

\noindent \lbrack 36] Ivezi\'{c} T 2012 \textit{Preprint }1212.4684

\noindent \lbrack 37] Ivezi\'{c} T 2015 \textit{Preprint }1508.04802

\noindent \lbrack 38] Oziewicz Z 2011 \textit{J. Phys.: Conf. Ser.} \textbf{%
330} 012012

\noindent \lbrack 39] Ivezi\'{c} T 2008 \textit{Fizika A} \textbf{17} 1

\end{document}